\documentclass{aa}  

\usepackage{graphicx}   
\usepackage{txfonts}    

\usepackage{natbib}
\bibpunct{(}{)}{;}{a}{}{,} 

\newcommand{\ferre}{FER\reflectbox{R}E} 
\newcommand{\gaia}{\textit{Gaia} }      
\newcommand{\kms}{\,km\,s$^{-1}$}
\newcommand{\minus}{\scalebox{0.75}[1.0]{$-$}}

\DeclareRobustCommand{\teff}{$T_{\mathrm{eff}}$}
\DeclareRobustCommand{\logg}{$\log g$}
\DeclareRobustCommand{\mh}{$\mathrm{[M/H]}$}
\DeclareRobustCommand{\vmic}{$v_{\mathrm{mic}}$}

\begin{document}

   \title{One Star to Tag Them All (OSTTA):}

   \subtitle{I. Radial velocities and chemical abundances for 20 poorly studied open clusters\footnote{The full versions of Tables \ref{tab:tableAP} and \ref{{tab:avg_abu}} are only available in electronic form
at the CDS via anonymous ftp to cdsarc.u-strasbg.fr (130.79.128.5)
or via http://cdsweb.u-strasbg.fr/cgi-bin/qcat?J/A+A/.}}

   \author{R. Carrera\inst{1}
          \and
          L. Casamiquela\inst{2}
          \and
          A. Bragaglia\inst{3}
          \and
          E. Carretta\inst{3}
          \and
          J. Carbajo-Hijarrubia\inst{4}
          \and
          C. Jordi\inst{4}
          \and
          J. Alonso-Santiago\inst{5}
          \and
          L. Balaguer-Nu\~nez\inst{4}
          \and
          M. Baratella\inst{6}
          \and
          V. D'Orazi\inst{1}
          \and
          S. Lucatello\inst{1}
          \and
          C. Soubiran\inst{2}
          }

   \institute{INAF-Osservatorio Astronomico di Padova, vicolo Osservatorio 5, 35122, Padova, Italy\\
              \email{jimenez.carrera@inaf.it}
         \and
             Laboratoire d’Astrophysique de Bordeaux, Univ. Bordeaux, CNRS, B18N, allée Geoffroy Saint-Hilaire, 33615 Pessac, France
        \and 
        INAF-Osservatorio di Astrofisica e Scienza dello Spazio di Bologna, via P. Gobetti 93/3, 40129, Bologna, Italy
        \and
        Institut de Ci\`encies del Cosmos, Universitat de Barcelona (IEEC-UB), Mart\'i i Franqu\`es 1, E-08028 Barcelona, Spain\\
        \and
        INAF–Osservatorio Astrofisico di Catania, via S. Sofia 78, 95123 Catania, Italy
        \and
        Leibniz-Institute for Astrophysics Potsdam (AIP), An der Sternwarte 16, D-14482, Potsdam, Germany
        }

   \date{Received September 15, 1996; accepted March 16, 1997}

  \abstract
   {Open clusters are ideal laboratories to investigate a variety of astrophysical topics, from the properties of the Galactic disc to stellar evolution models. For this purpose, we need to know their chemical composition in detail. Unfortunately, the number of systems with chemical abundances determined from high resolution spectroscopy remains small.}
   {Our aim is to increase the number of open clusters with radial velocities and chemical abundances determined from high resolution spectroscopy by sampling a few stars in clusters which had not been previously studied.}
   {We obtained high resolution spectra with the FIbre-fed Echelle Spectrograph (FIES) at Nordic Optical Telescope (NOT) for 41 stars belonging to 20 open clusters. These stars have high  astrometric membership probabilities determined from the \gaia second data release.}
   {We derived radial velocites for all the observed stars which were used to confirm their membership to the corresponding clusters. For Gulliver\,37, we cannot be sure the observed star is a real member. We derived atmospheric parameters for the 32 stars considered to be real cluster members. We discarded five stars because they have very low gravity or their atmospheric parameters were not properly constrained due to low signal-to-noise ratio spectra. Therefore, detailed chemical abundances were determined for 28 stars belonging to 17 clusters. For most of them, this is the first chemical analysis available in the literature. Finally, we compared the clusters in our sample to a large population of well-studied clusters. The studied systems follow the trends, both chemical and kinematical, described by the majority of open clusters. It is worth mentioning that the three most metal-poor studied clusters (NGC\,1027, NGC\,1750, and Trumpler\,2) are enhanced in Si, but not in the other $\alpha$-elements studied (Mg, Ca, and Ti).}
   {}


   \maketitle
%

\section{Introduction}

The \textit{Gaia} mission \citep{gaia2016mission} is providing unprecedented accurate positions ($\alpha$ and $\delta$), proper motions ($\mu_{\alpha}$ and $\mu_{\delta}$), and parallaxes ($\varpi$) for more than 1.8 billion stars \citep{gaia_edr3brown}. Additionally, \gaia also measures magnitudes in three photometric bands, \textit{G, $G_{BP}$,} and \textit{$G_{RP}$,} providing a unique homogeneous database \citep{gaia_edr3riello}.
Although in a more limited way \gaia also provides radial velocities, presently they are available for more than 7 million stars with \textit{$G_{RP}$}$\leq$13\,mag \citep{gaiadr2rv} but their number will greatly increase with further data releases. The radial velocities are derived with the Radial Velocity Spectrometer (RVS), working at a resolution of R=11\,500 and with small wavelength coverage centred at the infrared \mbox{Ca\,{\sc ii}} triplet, 845-872\,nm \citep{gaiadr2rvs}. Starting from \gaia data release 3\footnote{https://www.cosmos.esa.int/web/gaia/dr3}, metallicities and some abundances will also be provided based on RVS data. While \gaia RVS will provide the largest spectroscopic data set ever, there are also many limitations, both in magnitude limits and in the precision reached. This is why different follow-up intermediate resolution (e.g. R$\sim$20\,000) spectroscopic surveys have been organised in order to complement the \gaia capabilities, such as the Gaia-ESO Survey \citep[GES,][]{ges_gilmore,ges_randich}, Apache Point Observatory Galactic Evolution Experiment \citep[APOGEE,][]{apogee}, GALactic Archaeology with HERMES \citep[GALAH,][]{desilva2015_galah}, the forthcoming WHT Enhanced Area Velocity Explorer \citep[WEAVE,][]{weave2012}, and 4-metre Multi-Object Spectroscopic Telescope \citep[4MOST,][]{4most} Galactic surveys. In total, they are going to measure radial velocities and chemical abundances for a few million stars. All together, they are leading to a revolution in  our  knowledge  of the Milky Way as well as the surrounding galaxies including Galactic stellar clusters, of course. 

Stellar clusters are groups of stars sharing the same general properties (age, distance, and initial chemical composition), and they are the ideal laboratories to test models of stellar and Galactic formation and evolution \citep{freemanaraa2002}. In particular, open clusters (OCs) can trace the disc properties
and, when carefully selected, they can be used as observational constraints for stellar models from low to high mass, having ages from a few million years to about 10\,Ga \citep[e.g.][]{bragaglia_tosi2006bocce,cantatgaudin2020,dias2021}. 
\gaia high precision astrometry has allowed known open
clusters to be confirmed or discarded \citep[e.g.][]{Cantat-Gaudin+2018} and it has led to the discovery of hundreds of new ones \citep[e.g.][]{Castro-Ginard+2018,Castro-Ginard+2019,Castro-Ginard+2020}. A full characterisation and an accurate age derivation requires, however, one to know
the metallicity and the full set of detailed abundances from all nucleosynthetic chains as well. For example, \citet{bossini2019} determined ages for almost 270 OCs from \gaia photometry finding, on the other hand, degeneracies in almost all cases with metallicity.

Unfortunately, more than 90\% of the about 3000 known OCs have never been studied using high resolution spectroscopy \citep[e.g.][]{donati2015tr5,netopil2016OChomogeneous,netopil2022,Magrini+2017,occaso2,smiljanic2018_AA616}.
This paucity is currently or will be partially filled by the named intermediate resolution surveys mentioned before. For example, APOGEE has observed stars belonging to about 130 OCs \citep{carrera2019apo,donor2020dr16} and GES has also done so for those belonging to about 80 clusters \citep[][Randich et al. {\sl in prep.}]{bragaglia2021}. The
key feature of these surveys resides in their ability to sample numerous clusters or in studying a significant fraction of the clusters members in all
evolutionary phases, with tens to many hundred of stars observed in each cluster, as GES did. This is important to
understand both the formation of the clusters \citep[e.g.][]{jeffries2014,mapelli2015} and the evolution of the stars’ properties depending on rotation, activity, surface
abundances, and key constraints to modern stellar evolutionary models \citep[e.g.][]{BertelliMotta+2018,Smiljanic+2016}. However, spectra of stars with
the high metallicity typical of OCs are difficult to analyse at the intermediate resolution common to
all large surveys because of line blending\footnote{Indeed, GES uses the high resolution UVES spectra for complete 
chemical characterisation.}. Only spectra with simultaneous high resolution (R$>$50\,000) and wide spectral coverage can allow the measurements
of the full set of the Fe-peak, CNO, $\alpha$-elements, neutron-capture elements, etc., with the necessary high precision and accuracy.

The One Star  to Tag Them All (OSTTA) project was designed to provide a high resolution spectroscopy follow-up of poorly studied open clusters. To do that, we have acquired spectra of a few objects per cluster in order to determine radial velocities and measure abundances of elements synthesised through all nucleosynthesis chains, therefore providing robust constraints
to stellar evolutionary models and to the history of the Galactic disc. In this paper, we present the results for 41 stars in 20 open clusters.

The paper is organised as follows. The target selection, observations, and data reduction are discussed in Sect.~\ref{sec:obs}. The radial velocity determination is described in Sect.~\ref{sec:rv}. The chemical abundance determination is presented in Sect.~\ref{sec:abun_determinatiom}. The results are discussed in the context of the majority of Galactic OCs in Sect.~\ref{sec:discussion} including the orbit determination. Finally, the conclusions are summarised in Sect.~\ref{sec:conclusion}.

\section{Target selection, observations, and data reduction}\label{sec:obs}

\begin{table*}
\setlength{\tabcolsep}{1.25mm}
\begin{center}
\caption{Parameters of the observed clusters.}
\begin{tabular}{lcccccccccccc}
\hline
Cluster & $\alpha_{\rm ICRS}$ & $\delta_{\rm ICRS}$ & $\varpi$ & $\mu_{\alpha}$ & $\mu_{\delta}$ & Age & $A_{\rm V}$ & Distance & $X$ & $Y$ & $Z$ & $R_{\rm GC}$\\
    & [deg]    & [deg] & [mas] & [mas~yr$^{-1})$] & [mas~yr$^{-1}$]  & [Ga] & [mag] & [pc] & [pc] & [pc] & [pc] & [pc]\\
\hline
  ASCC\,23 & 95.047 & 46.71 & 1.59$\pm$0.06 & 1.10$\pm$0.18 & -0.60$\pm$0.18 & 0.23 & 0.34 & 630 & -595 & 132 & 156 & 8936\\
  Alessi\,44 & 295.325 & 1.592 & 1.50$\pm$0.22 & 0.46$\pm$0.40 & -2.32$\pm$0.18 & 0.22 & 0.71 & 679 & 511 & 430 & -121 & 7840\\
  Alessi\,62 & 284.026 & 21.597 & 1.59$\pm$0.06 & 0.24$\pm$0.16 & -1.07$\pm$0.18 & 0.69 & 0.78 & 650 & 388 & 512 & 98 & 7967\\
  COIN-Gaia\,2 & 15.06 & 55.409 & 0.79$\pm$0.06 & -4.46$\pm$0.12 & -1.93$\pm$0.12 & 0.66 & 0.46 & 1204 & -670 & 987 & -155 & 9064\\
  COIN-Gaia\,6 & 28.101 & 58.636 & 0.28$\pm$0.06 & -2.35$\pm$0.11 & -0.49$\pm$0.14 & 0.48 & 1.18 & 3259 & -2126 & 2463 & -187 & 10752\\
  FSR\,0951 & 95.573 & 14.65 & 0.55$\pm$0.05 & 0.24$\pm$0.12 & 0.07$\pm$0.10 & 0.52 & 0.81 & 1757 & -1684 & -502 & 10 & 10036\\
  Gulliver\,37 & 292.077 & 25.347 & 0.64$\pm$0.04 & -0.77$\pm$0.07 & -3.74$\pm$0.09 & 0.35 & 1.33 & 1438 & 727 & 1237 & 95 & 7712\\
  King\,23 & 110.459 & -0.988 & 0.28$\pm$0.06 & -0.46$\pm$0.08 & -0.88$\pm$0.09 & 1.95 & 0.17 & 3136 & -2480 & -1889 & 344 & 10983\\
  NGC\,581 & 23.339 & 60.659 & 0.37$\pm$0.04 & -1.38$\pm$0.071 & -0.50$\pm$0.08 & 0.03 & 1.09 & 2502 & -1541 & 1969 & -78 & 10075\\
  NGC\,1027 & 40.677 & 61.616 & 0.88$\pm$0.05 & -1.75$\pm$0.13 & 2.09$\pm$0.17 & 0.19 & 1.01 & 1125 & -805 & 785 & 30 & 9179\\
  NGC\,1647 & 71.481 & 19.079 & 1.67$\pm$0.08 & -1.06$\pm$0.25 & -1.50$\pm$0.24 & 0.36 & 0.64 & 635 & -608 & -3 & -183 & 8948\\
  NGC\,1750 & 75.926 & 23.695 & 1.36$\pm$0.09 & -0.96$\pm$0.25 & -2.37$\pm$0.20 & 0.26 & 0.79 & 727 & -715 & 10 & -135 & 9055\\
  NGC\,2186 & 93.031 & 5.453 & 0.41$\pm$0.05 & 0.41$\pm$0.12 & -1.99$\pm$0.11 & 0.25 & 0.65 & 2251 & -2052 & -894 & -243 & 10430\\
  NGC\,2281 & 102.091 & 41.06 & 1.90$\pm$0.09 & -2.95$\pm$0.27 & -8.32$\pm$0.26 & 0.62 & 0.09 & 544 & -518 & 46 & 158 & 8858\\
  NGC\,2345 & 107.075 & -13.199 & 0.35$\pm$0.05 & -1.33$\pm$0.10 & 1.34$\pm$0.11 & 0.21 & 1.04 & 2663 & -1829 & -1933 & -107 & 10351\\
  NGC\,2358 & 109.261 & -17.143 & 1.06$\pm$0.05 & -1.35$\pm$0.08 & 0.54$\pm$0.09 & 0.66 & 0.02 & 908 & -570 & -705 & -35 & 8938\\
  NGC\,7654 & 351.195 & 61.59 & 0.60$\pm$0.05 & -1.94$\pm$0.15 & -1.13$\pm$0.15 & 0.15 & 1.85 & 1653 & -641 & 1524 & 12 & 9109\\
  Stock\,1 & 294.146 & 25.163 & 2.45$\pm$0.08 & 6.03$\pm$0.31 & 0.30$\pm$0.33 & 0.42 & 0.39 & 416 & 206 & 361 & 15 & 8141\\
  Trumpler\,2 & 39.232 & 55.905 & 1.43$\pm$0.06 & 1.58$\pm$0.19 & -5.35$\pm$0.17 & 0.11 & 0.86 & 710 & -521 & 479 & -49 & 8874\\
  UBC\,54 & 64.747 & 46.453 & 0.88$\pm$0.05 & 3.32$\pm$0.11 & -3.76$\pm$0.11 & 0.25 & 0.98 & 1105 & -1006 & 452 & -52 & 9357\\
\hline
\end{tabular}
\tablefoot{
Values obtained from \citet{cantatgaudin2020} derived from \gaia DR2, except for proper motions and parallaxes, which have been recomputed using \gaia EDR3.
}
\label{tab:ocs}
\end{center}
\end{table*}

\begin{figure*}
   \centering
   \includegraphics[width=\textwidth]{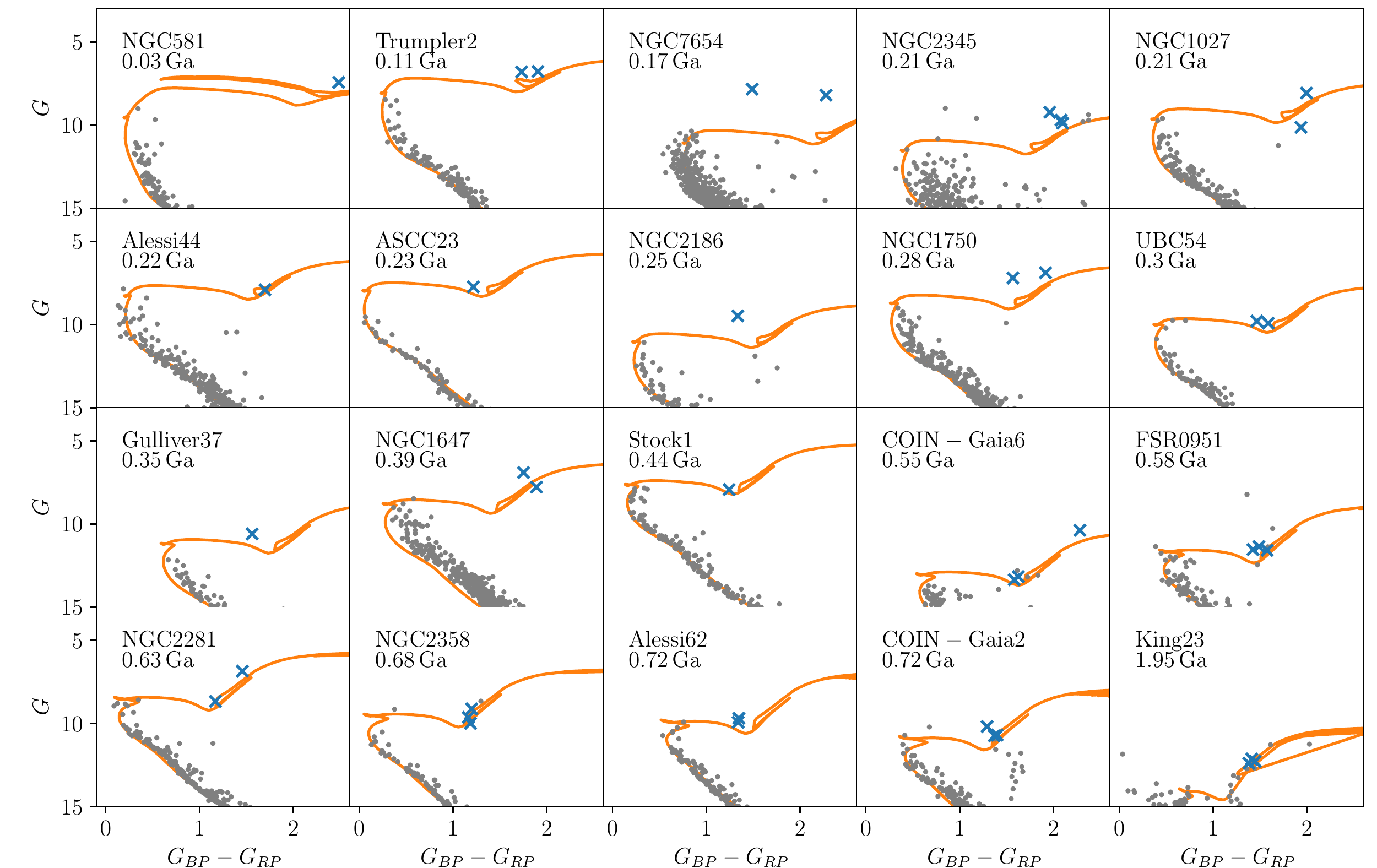}
   \caption{\gaia DR2 colour-magnitude diagrams of the observed clusters, taking member stars from \citet{cantatgaudin2020} (grey points). Blue crosses are spectroscopic targets sampled here. PARSEC isochrones \citep{Marigo+2017} at the metallicity computed in this paper were overplotted using distances, ages, and extinctions provided by \citet{cantatgaudin2020}. The clusters are ordered by increasing age. Ages are labelled in each panel.}
   \label{fig:dcms}%
\end{figure*}

The target selection was based on the astrometric membership probabilities determined by \citet{cantatgaudin2018} for 1229 clusters taking advantage of the \gaia data release 2 (DR2) positions, proper motions, and parallaxes. These membership probabilities were determined by the unsupervised photometric membership assignment in stellar clusters (UPMASK) which is based on the $\kappa$-means clustering algorithm. We refer the reader to \citet{cantatgaudin2018} for details on how the probabilities were assigned. We selected OCs for which composition based on spectroscopy was not available in the literature at the time of observations, and where at least a giant star brighter than $G\sim$14\,mag is among the most probable members. We also included stars in cluster UBC\,54 which were first reported by \citet{Castro-Ginard+2018} on the basis of \gaia DR2 positions, proper motions, and parallaxes. The list of sampled clusters together with their main features are listed in Table~\ref{tab:ocs}. The observed stars are listed in Table~\ref{tab:obs_log} and their position in the colour-magnitude diagram of each cluster is shown in Fig.~\ref{fig:dcms}.

The observations were acquired during five nights, 12-16 December 2018, and complemented in another three nights, 18,19, and 24 April 2019, with the FIbre-fed Echelle Spectrograph \citep[FIES,][]{fies} in its highest spectral resolution, R$\sim$67\,000. FIES is installed at the 2.5\,m Nordic Optical
Telescope (NOT) at the Observatorio del Roque de los Muchachos on the island of La Palma (Spain). The number of exposures for each star, together with the final signal-to-noise ratio (S/N), total exposure times, and dates of observations are summarised in Table~\ref{tab:obs_log}. 

The bias subtraction, flat-field correction, order tracing, and extraction and wavelength calibration were performed by the FIEStool dedicated pipeline specifically developed for this instrument. After this, the sky and telluric subtraction were performed from the extracted and wavelength calibrated spectra still separated by orders. Then, we proceeded with the combination, normalisation, and merge of the different orders. All the steps were performed with the tools developed in the framework of the OCCASO\footnote{Open Clusters Chemical Abundances from Spanish Observatories survey \citep[][]{occaso1}.} survey and described in detail by \citet{occaso4}.

\section{Radial velocities}\label{sec:rv}

Radial velocities, $v_{\rm rad}$, were measured as in \citet{occaso4}. Briefly, the radial velocities of the 1D combined spectra were obtained by measuring the Doppler shifts of the spectral lines using the classical cross-correlation method \citep[e.g][]{tonrydavis}. To do that, the observed spectrum was cross-correlated against a template synthetic spectrum which better reproduces the observed one. The synthetic template for each star was selected from a synthetic grid using \ferre\footnote{Available at \url{https://github.com/callendeprieto}} \citep{allende_ferre}. We used the coarse {\sl hnsc1}, {\sl hnsc2}, and {\sl hnsc3} grids described by \citet{allendegrids} with the following three dimensions: metallicity [M/H]; effective temperature, T$_{\rm eff}$; and surface gravity, $\log g$. All together, these grids cover spectral types between early M ($T_{\rm eff}$=3\,500\,K) and A ($T_{\rm eff}$=12\,000\,K). We refer the reader to \citet{allendegrids} for details about the ranges of the parameters covered by each grid. The grids, originally computed with an infinity resolution and 0.45\,\kms sampling equivalent to a resolution of 300\,000, were smoothed to match the nominal resolution of FIES of 67\,000.

Our procedure provides two independent determinations of the radial velocity uncertainties. The first one, $v_{\rm err}$, is determined from the height of the cross-correlation peak. This value mainly depends on how well the template spectrum reproduces the observed one. The second one, $v_{\rm scatter}$, is calculated as the velocity scatter of the individual measurements for each star, taking advantage of the fact that, in most cases, at least three individual exposures have been collected for each object. This value was obtained at the moment of combining the individual exposures by cross-matching each of them with the combined one, and therefore, it is model-independent. In principle, $v_{\rm scatter}$ should be a better estimation of our internal precision than $v_{\rm err}$, which systematically underestimates the radial velocity uncertainty. Moreover, large values of $v_{\rm scatter}$ can be used as an indicator of binarity if individual exposures are adequately separated in time. Unfortunately, in our case, the individual exposures were often taken one after the other. From the detailed analysis presented by \citet{occaso4}, the typical $v_{\rm scatter}$ value for the instrumental configuration used in our case is 15.4\,m\,s$^{\rm -1}$.

The obtained radial velocities together with their uncertainties, both $v_{\rm scatter}$ and $v_{\rm err}$, and the number of individual exposures used in each case are listed in Table~\ref{tab:obs_log}. We used exactly the same tools as in \citet{occaso4}, who have compared the larger OCCASO sample with different spectroscopic surveys available with the literature. They found that the obtained values are in agreement with the literature ones within the uncertainties involved in each case. We refer the reader to this paper for details. In this paper, we are going to concentrate on the individual comparison of our radial velocity with previous determinations available in the literature.

\begin{figure}
\centering
\includegraphics[width=\columnwidth]{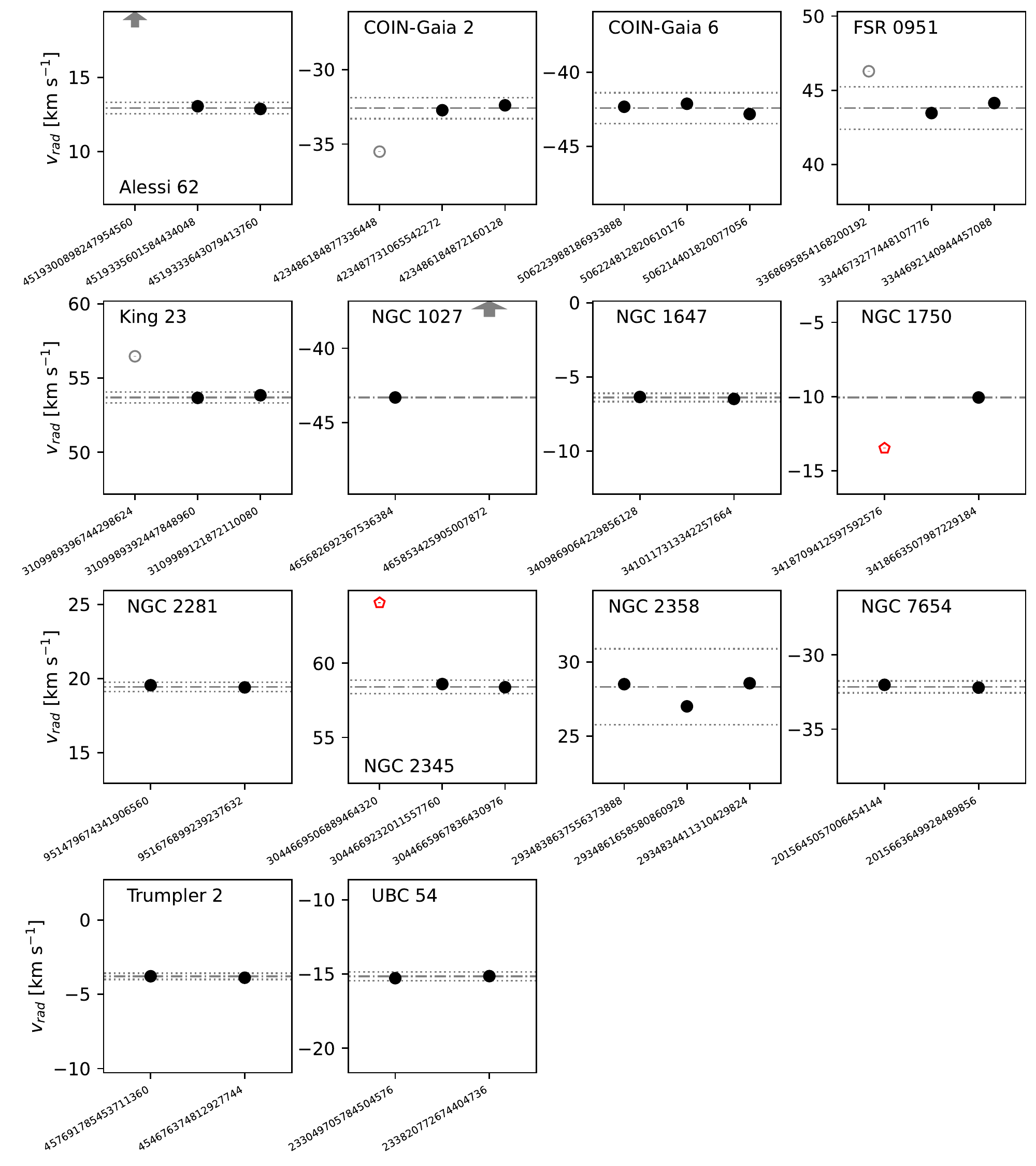}
\caption{Determination of the average radial velocity for each cluster with two or more stars sampled. Filled symbols mark stars considered real cluster members, while open symbols are the excluded objects. Red pentagons are the reported spectroscopic binaries. Arrows denote the objects outside the panels. Dot-dashed lines correspond to the average radial velocity for each cluster. Dotted lines show the $3\times\sigma_{v_{\rm rad}}$ level. We note that the error bars are smaller than the symbol size.}
\label{fig:rv_membership}%
\end{figure}

For those clusters where two or more stars were observed, the average radial velocity was derived following the equations described by \citet{soubiran2018gaiadr2_rv} and it is shown in Fig.~\ref{fig:rv_membership}. The small number of stars sampled in each cluster -- for which there are three in the best case -- prevented us from using statistical techniques, such as an iterative $\kappa$-$\sigma$ clipping algorithm, in order to remove objects with discrepant radial velocities. Therefore, the membership probability of each star and the reliability of the average cluster radial velocity is discussed on the basis of the astrometric membership probabilities, $p$, determined from \gaia DR2 by \citet{cantatgaudin2018} and \citet{cantatgaudin2020}; the radial velocities derived here; and other determinations available in the literature by \citet[][S18]{soubiran2018gaiadr2_rv} from \gaia DR2 radial velocities, \citet[][T21]{tarricq2021} from a compilation including also \gaia DR2, and \citet[][M08]{Mermilliod+2008} from their own observations.

\begin{table*}[h!]
\setlength{\tabcolsep}{1mm}
\begin{center}
\caption{Mean radial velocities obtained here for the observed clusters, together with other values available in the literature.}
\begin{tabular}{lcccccccccccccc}
\hline
Cluster & $v_{\rm rad}$ & $\sigma_{v_{\rm rad}}$ & $e_{v_{\rm rad}}$\tablefootmark{a} & N & $v_{\rm S18}$ & $e_{v_{\rm S18}}$ & $\sigma_{v_{\rm S18}}$ & N$_{\rm S18}$ & $v_{\rm T21}$ & $e_{v_{\rm T21}}$ & N$_{T21}$ & $v_{\rm M08}$ & $\sigma_{\rm M08}$ & N$_{\rm M08}$ \\
\hline
  ASCC\,23 & -13.35 &  & 0.02 & 1 & -13.29 & 1.49 & 3.65 & 6 & -13.3 & 1.38 & 9 &  &  &  \\
  Alessi\,44 & -9.58 &  & 1.11 & 1 & -9.8 & 0.67 & 1.78 & 7 & -10.68 & 12.95 & 9 &  &  &  \\
  Alessi\,62 & 12.93 & 0.13 & 0.09 & 2 & 13.28 & 0.2 & 0.68 & 12 & 13.28 & 0.21 & 10 &  &  &  \\
  COIN-Gaia\,2 & -32.58 & 0.23 & 0.17 & 2 &  &  &  &  & -34.55 & 2.41 & 3 &  &  &  \\
  COIN-Gaia\,6 & -42.41 & 0.35 & 0.20 & 3 &  &  &  &  & -47.26 & 3.75 & 3 &  &  &  \\
  FSR\,0951 & 43.80 & 0.48 & 0.34 & 2 & 45.53 & 0.55 & 1.24 & 5 & 45.01 & 2.04 & 5 &  &  &  \\
  King\,23 & 53.69 & 0.12 & 0.08 & 2 & 54.01 & 0.43 & 0.86 & 4 & 54.01 & 0.43 & 4 &  &  &  \\
  NGC\,581 & -46.78 &  & 0.01 & 1 & -45.33 & 0.32 & 0.0 & 1 & -44.09 & 0.7 & 1 & -44.2 & 0.12 & 2 \\
  NGC\,1027 & -43.32 & & 0.02 & 1 & -4.06 & 0.31 & 1 & -36.57 & 10.83 & 6 & & & \\
  NGC\,1647 & -6.39 & 0.09 & 0.06 & 2 & -6.69 & 0.21 & 0.86 & 16 & -6.71 & 0.17 & 21 & -7.02 & 0.22 & 2 \\
  NGC\,1750 & -10.06 &  & 0.01 & 1 & -10.38 & 0.87 & 2.31 & 7 & -7.45 & 6.64 & 13 &  &  &  \\
  NGC\,2186 & 20.82 &  & 0.02 & 1 & 21.98 & 0.26 & 0.0 & 1 & 21.71 & 4.01 & 3 & 20.15 & 0.19 & 1 \\
  NGC\,2281 & 19.44 & 0.10 & 0.07 & 2 & 19.58 & 0.34 & 2.19 & 42 & 18.95 & 0.95 & 40 & 19.05 & 0.04 & 2 \\
  NGC\,2345 & 58.41 & 0.15 & 0.11 & 2 & 60.94 & 2.0 & 4.89 & 6 & 63.27 & 2.43 & 6 & 59.19 & 0.36 & 4 \\
  NGC\,2358 & 28.31 & 0.85 & 0.49 & 3 & 27.57 & 0.6 & 1.03 & 3 & 27.57 & 0.6 & 3 &  &  &  \\
  NGC\,7654 & -32.15 & 0.13 & 0.09 & 2 & -32.09 & 0.13 & 0.1 & 2 & -32.18 & 0.27 & 2 & -32.98 & 0.11 & 1 \\
  Stock\,1 & -19.60 &  & 0.24 & 1 & -19.51 & 0.53 & 2.91 & 30 & -19.52 & 0.55 & 29 &  &  &  \\
  Trumpler\,2 & -3.78 & 0.07 & 0.05 & 2 & -4.12 & 0.09 & 0.28 & 11 & -3.97 & 1.14 & 12 &  &  &  \\
  UBC\,54 & -15.15 & 0.10 & 0.07 & 2 &  &  &  &  & -15.26 & 0.33 & 2 &  &  &  \\
\hline
\end{tabular}
\tablefoot{\tablefoottext{a}{For those clusters with a single observed star, $e_{v_{\rm rad}}$ is the $v_{\rm scatter}$ of that object.}}

\label{tab:oc_avglit}
\end{center}
\end{table*}

A detailed discussion of each particular cluster can be found in appendix~\ref{apex:indvnotes}. The final values are listed in Table~\ref{tab:oc_avglit} together with other values available in the literature for comparison. A total of six stars were discarded as members by comparison with other objects in the same system for clusters Alessi\,62, COIN-Gaia\,2, FSR\,0951, King\,23, NGC\,1027, and NGC\,1750. Moreover, we considered another three stars as spectroscopic binaries in clusters Gulliver\,37 and NGC\,2345. Therefore, we removed Gulliver\,37 from our sample since we cannot ensure that the only star sampled in this system is a real cluster member. These objects are properly flagged in Table~\ref{tab:obs_log}.

\section{Chemical abundance determination}\label{sec:abun_determinatiom}

We derived atmospheric parameters and individual chemical abundances for all the stars considered as real cluster members in the previous section. We excluded the eight stars considered as non-members or spectroscopic binaries in the previous section.

\subsection{Method}
The atmospheric parameters (\teff, \logg, \mh) and individual chemical abundances were derived using the public spectroscopic software iSpec \citep{BlancoCuaresma+2014,BlancoCuaresma2019}. We made use of the synthetic spectral synthesis method in the 1D Local Thermodynamic Equilibrium (LTE) approach, with the radiative transfer code SPECTRUM \citep{Gray+1994} and the MARCS atmospheric models \citep{Gustafsson2008}. We used the latest version (the sixth one) of the line list from GES \citep{Heiter+2021}. Atmospheric parameters and abundances were obtained in a two-step process, similarly as what was done in \citet{Casamiquela+2020}.

The atmospheric parameters \teff, \logg, \mh, and $[\alpha\mathrm{/M}]$, as well as the microturbulence parameter \vmic\ were inferred for each spectrum using spectral synthesis fitting. The selection of lines from the master line list is the same as the one used in \citet{occaso2}, which includes a sample of 330 atomic lines from different elements selected to be sensitive to the atmospheric parameters at the FIES resolution (67\,000). In this step, we also used the wings of H$\alpha$ and H$\beta$, as well as the \mbox{Mg\,{\sc i}} \textit{b} triplet lines, which are sensitive to the \teff. As for the broadening parameters, we set the same strategy as in \citet{Casamiquela+2020}: only the spectral resolution was let free throughout the iterations, accounting for all broadening effects. We fixed the atmospheric parameters to the results of the previous step to determine the absolute chemical abundances of individual lines, also using spectral synthesis fitting.

\begin{figure}
\centering
   \includegraphics[width=0.5\textwidth]{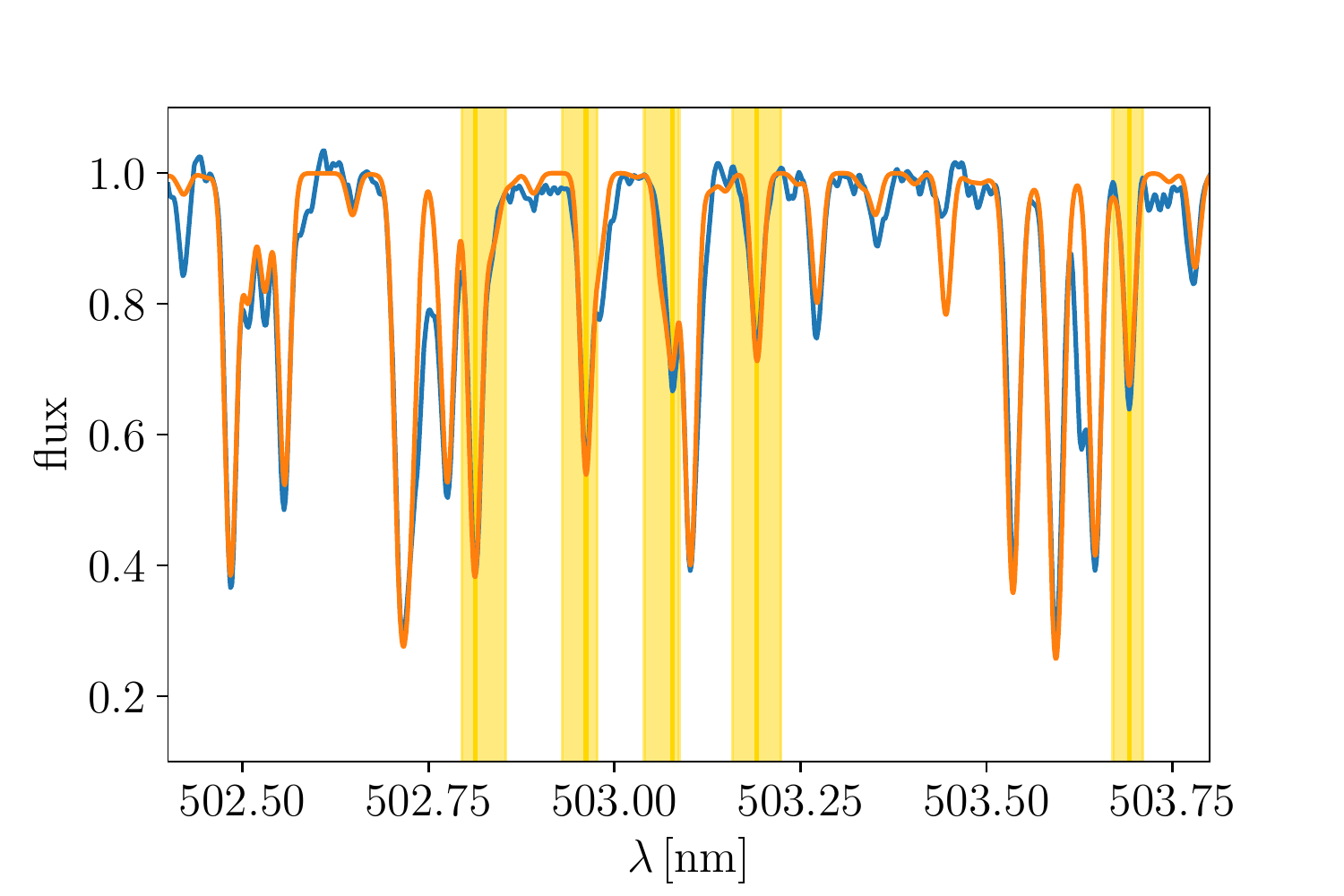}
   \caption{Portion of the observed spectrum of the star \gaia EDR3 2934861658580860928 (blue) and the best synthetic fit (orange) obtained to constrain the atmospheric parameters. The line masks where the fit was performed are marked in yellow, and they correspond to five \mbox{Fe\,{\sc i}} lines.}\label{fig:spectrum}
\end{figure}

The spectral fitting was done by comparing the observed fluxes weighted by their uncertainties with a synthetic spectrum for the mask regions of the selected spectral features. The best fit was selected using a least-squares algorithm. The uncertainty in the atmospheric parameters was obtained from the $\chi^2$ of the fit. Figure~\ref{fig:spectrum} shows an example of the observed spectrum for star \gaia   Early data release 3 (EDR3) 2934861658580860928 of the cluster NGC\,2358 with several selected features and the corresponding fits.

For the individual abundance determination, we used a custom line selection built in the same way as in \citet{Casamiquela+2020} but performing with the spectra used in this study. In brief, it consisted in keeping the lines that give consistent abundances in all stars for elements with a high enough number of detected initial lines ($\gtrsim10$). For the elements with few lines available, we kept the lines based on the flags provided by the GES line list group (LOGGFFLAG and SYNFLAG). Our final selection contained 209 lines from 17 chemical species.

\subsection{Atmospheric parameters}\label{sect:atmos_dete}

\begin{figure*}
\centering
   \includegraphics[width=\textwidth]{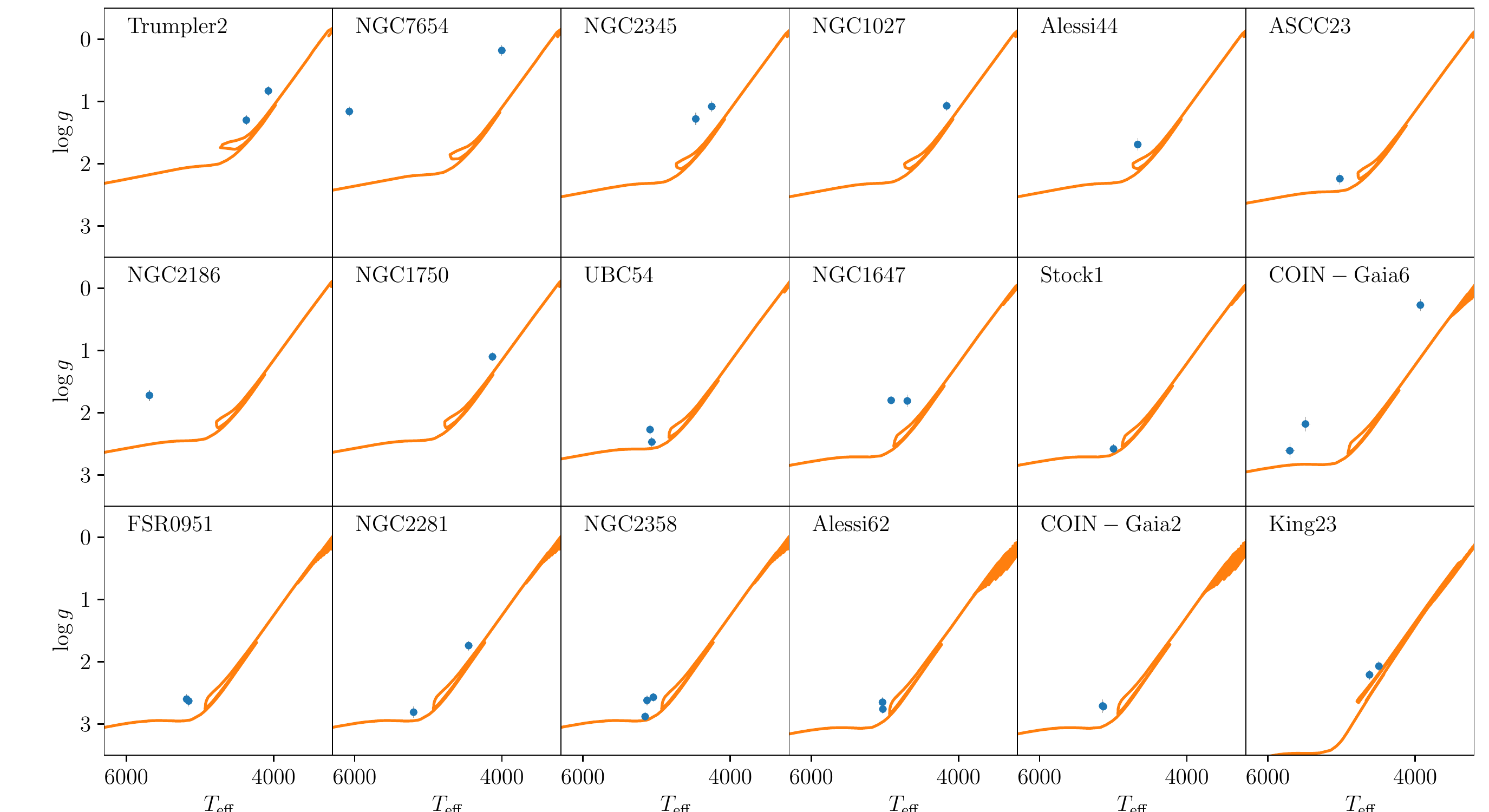}
   \caption{Kiel diagram of the analysed stars. As in Fig.~\ref{fig:dcms}, the corresponding isochrones for the age of each cluster are overplotted as a reference.}\label{fig:kiel}
\end{figure*}

The determined atmospheric parameters are listed in Table~\ref{tab:stars_info} and represented in Kiel diagrams in Fig.~\ref{fig:kiel}. Most of the stars belong to the red clump or the red giant branch, but there are several stars which are probably red supergiants, particularly one star of NGC\,7654 and another one of COIN-Gaia\,6 with surface gravities close to zero. Since their spectra are very crowded with atomic and molecular lines, this makes their determination of atmospheric parameters and abundances more uncertain. The other two stars in COIN-Gaia\,6 have low S/N spectra. Their spectra appeared very crowded, and the fits performed by our pipeline are not satisfactory. Finally, we also excluded the star observed in NGC\,581 which was reported as an M supergiant in the literature by \citet{keenan1989}. We tried to derive the atmospheric parameters for this star, but the obtained fit was of a very poor quality.
We excluded these suspected objects from our posterior analysis. 

Several stars appear far from the predicted isochrone: the remaining star in NGC\,7654 and the one in NGC\,2186. For these stars, the synthetic fits were satisfactory, and our pipeline converged well. In these two cases, we  also saw a shift from the isochrone in the colour-magnitude diagram (Fig.~\ref{fig:dcms}), thus probably the age of these clusters is not correct (see discussion about NGC\,7654 in Appendix~\ref{apex:indvnotes}).  

Finally, there are several cases (NGC\,1027, NGC\,1750, and NGC\,1647) for which stars appear far from the isochrone in the colour-magnitude diagrams, but not in the Kiel diagram. This thus indicates that  there is probably an issue in the photometry, such as an unresolved companion or maybe the colour-magnitude diagrams are affected by differential reddening.

NGC\,2345 is the only cluster in our sample which has been properly studied before. The same three stars sampled here were studied by \cite{Reddy+2016}, \citet{alonso_santiago2019}, and \citet{holanda2019}. One of them is a spectroscopic binary, as specified in Appendix~\ref{apex:indvnotes}. The results obtained here and in the literature are summarised in Table~\ref{tab:ngc2345_lit}. In general, there is good agreement within the uncertainties for the two stars. The exception is the gravity for the coolest star (3044665967836430976) for which \citet{Reddy+2016} report a significantly larger gravity. For the iron abundances, the values obtained here are higher by about 0.17\,dex, which is about twice that for the involved uncertainties. This could be explained by the existence of a zero-point between these measurements, although all of them used a similar Fe abundance for the Sun.

\citet{luck2014} determined atmospheric parameters for one of the stars in NGC\,7654 (2015663649928489856). The effective temperature (6114\,K) and metallicity (-0.06\,dex) are in agreement within the uncertainties with the values found here. The surface gravity determined by this author (1.63\,dex) is slightly higher than the value found here. Unfortunately, \citet{luck2014} does not provide information about the uncertainties of his determinations. Additionally, \citet{kovtyukh2007} also determined the effective temperature for the star 2015663649928489856 in NGC\,7654 (6268$\pm$149\,k), which again is in agreement with our value within the uncertainties. Additionally, the sampled star in Gulliver\,37 has been recently analysed by \citet{zhang_spa}. We excluded this object from our analysis because it was reported as a variable in the literature \citep{mermilliod2008_RVOC}. To our knowledge, these are the only determinations of atmospheric parameters available in the literature for the stars studied here.

\subsection{Chemical abundances}\label{sec:chemabs}

We computed abundances for the following 17 chemical species: Na, Mg, Al, Si, Ca, Sc, Ti, V, Cr, Mn, Fe, Co, Ni, Y, Ba, Ce, and Nd.
We derived our own Solar abundances from the analysis of a Solar spectrum of the spectral library of the \gaia FGK benchmark stars \citep{BlancoCuaresma+2014}, degraded to the FIES resolution. The determination of Solar abundances using the same analysis method allows one to partly take biases in the obtained [X/H] abundance values into account that could be caused by the performance of a particular method. The obtained solar abundances are listed in Table~\ref{tab:solar_abu}.

\begin{figure}
\centering
   \includegraphics[width=0.5\textwidth]{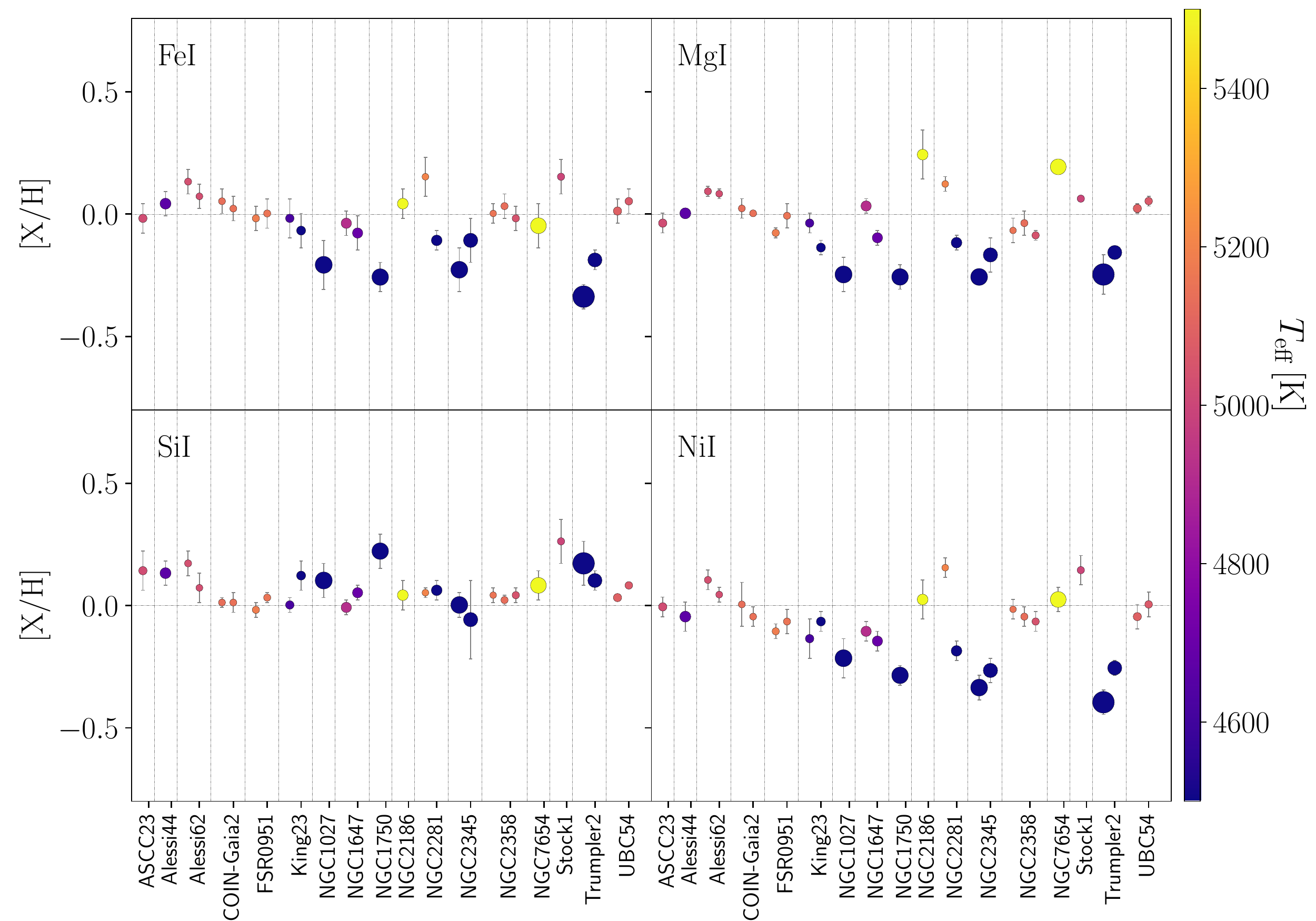}
   \caption{Individual star abundances from \mbox{Fe\,{\sc i}}, \mbox{Mg\,{\sc i}}, \mbox{Si\,{\sc i}}, and \mbox{Ni\,{\sc i}} lines. Stars are coloured according to the effective temperature, and their size is proportional to the surface gravity (larger points represent lower surface gravities). Vertical lines separate stars from the different clusters.}\label{fig:abus_ex}
\end{figure}

Individual stellar abundances were computed as the median of the individual line abundances, and the median absolute deviation is assigned as an error. When a single line was measured for a particular element, the error assigned is that of the line fit (computed from the residuals of the fit). As an example, we plotted in Fig.~\ref{fig:abus_ex} the Fe, Si, Mg, and Ni abundances for the sample of stars.

There exists a significant trend in most of the abundances with respect to the atmospheric parameters of the stars, except for Si. This behaviour is most probably caused by systematic effects from the analysis, and this has been thoroughly discussed in previous works \citep[see an extensive discussion in][]{Roederer+2014}. It can be caused for a variety of reasons, for example due to the fact that as temperature decreases lines can become more affected by blends, which are often not identified, and they can bias the abundance results. This bias can affect the various elements differently, as seen in \citet{Casamiquela+2020}, and it probably depends on the analysis method and line list used. A correction of such trends can be attempted by using empirical relations \citep[e.g.][]{Valenti+2005}, but it is usually difficult and risky because the underlying reasons are not always well understood. Moreover, to do so, a large number of stars with the same abundance and a wide coverage of the parameter space is needed, so we do not attempt to correct those trends in this study.

\begin{figure*}
\centering
   \includegraphics[width=\textwidth]{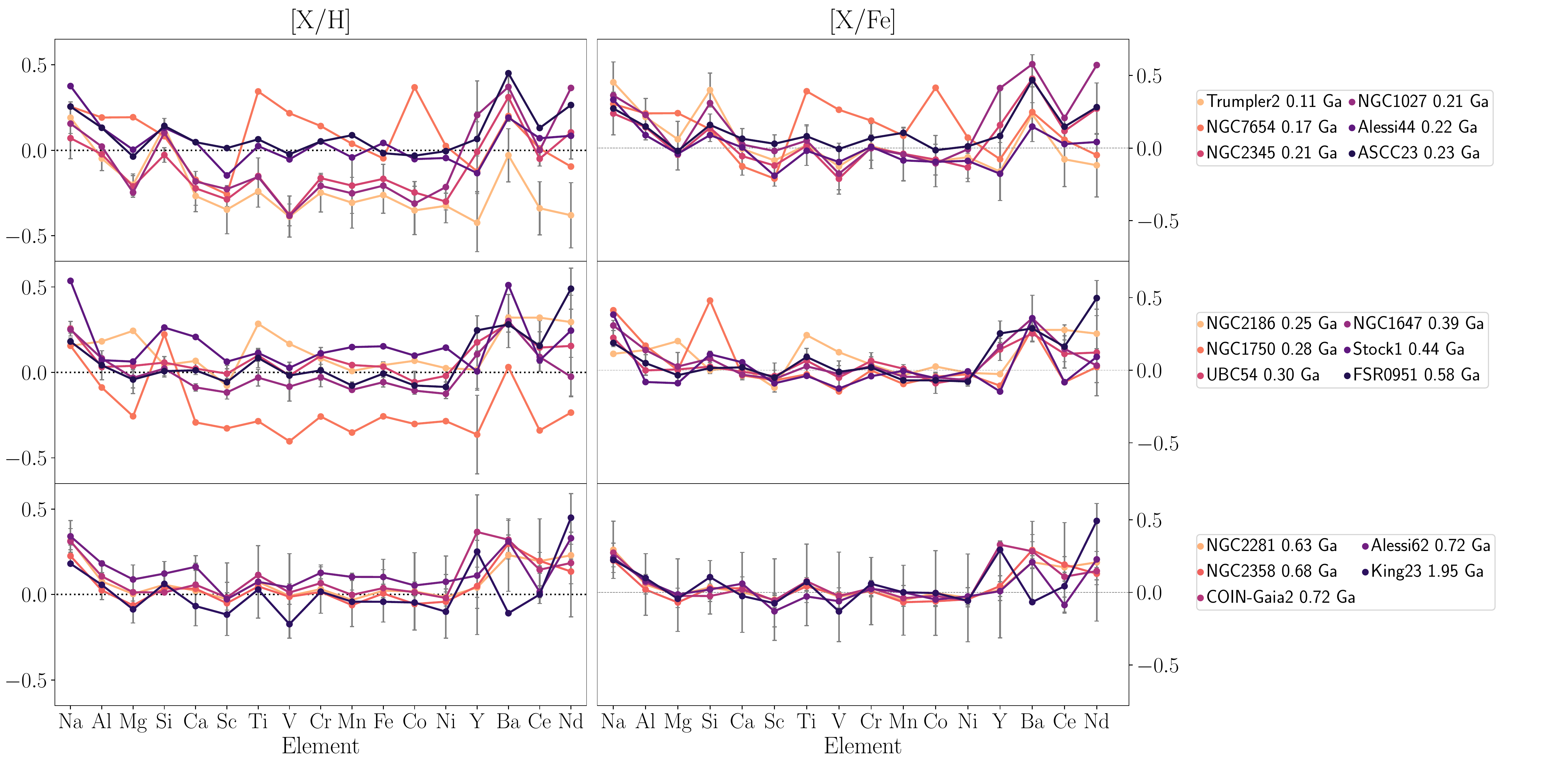}
   \caption{Mean cluster chemical abundances. Clusters are sorted by age in the panels increasing downwards, and in each panel from yellow (younger) to black (older).}\label{fig:abus}
\end{figure*}

In general, the studied clusters have nearly solar metallicities, as expected for their ages. The clusters that have more than one observed star (Alessi\,62, COIN-Gaia\,2, FSR\,0951, King\,23, NGC\,1647, NGC\,2281, NGC\,2345, NGC\,2358, Trumpler\,2, and UBC\,54) have good internal agreement in terms of chemical abundances. Trumpler\,2 is probably the only case where we see discrepant abundances at the level of 0.1\,dex among the two stars, remarkably in Ni and Fe, but not in Si and Mg. This can be explained by the difference in gravity among them of 0.5\,dex, which, as has been explained previously, can cause small biases in the abundances of certain elements.
Mean cluster abundances for all chemical species analysed are plotted in Fig.~\ref{fig:abus} and given in Table~\ref{tab:avg_abu}.

The clusters Trumpler\,2 and NGC\,1750 are the most metal-poor clusters in our sample, with iron abundances [Fe/H] of $-0.26\pm0.12$ and $-0.26\pm0.06$\,dex, respectively. This is surprising given their young age (110 and 280\,Ma, respectively). Similar cases have been recently reported in the literature \citep[e.g.][]{baratella2020,zhang_spa}, where clusters with relatively young ages show large discrepancies when compared to chemical models, with metallicities that are too
low for their Galactocentric positions. One of the hypotheses for this is the difficulty in deriving abundances for young stars due to their chromospheric activity and magnetic fields. A similar effect also explains the remarkable enhancement of Ba seen in our sample and in multiple works in the literature \citep[e.g.][]{Baratella+2021}. Alternatively, it could be that the lower Fe abundances obtained are partly explained by the previously mentioned trend of most elements with atmospheric parameters (see Fig.~\ref{fig:abus_ex}). Since the giant stars in younger clusters have smaller \teff\ and \logg, our abundances for the elements that exhibit a trend with atmospheric parameters could display a zero-point which depends on age.

Almost all clusters in our sample exhibit a Na enhancement. Similar enhancements have been found in the literature for red giants in clusters. High-mass giant stars can present overabundances up to 0.2\,dex \citep{Smiljanic+2016,smiljanic2018_AA616}, which are attributed to the effects of internal mixing and can be described by some stellar evolutionary models as in \citet{Lagarde+2012}.

\section{Observed clusters in the Galactic context}\label{sec:discussion}

In this section, we investigate the properties of the studied clusters in comparison with the global kinematics and chemical trends observed in the Galactic disc. 

\subsection{Open clusters' kinematics}

To check if the studied clusters follow the same kinematics as the majority of the OCs in the disc, we computed the line-of-sight velocity with respect to the Galactocentric standard of rest ($v_{\rm GSR}$) and with respect to the regional standard of rest\footnote{The regional standard of rest is defined as the local standard or rest at the position of each OC \citep{trumplerweaver1953}.} ($v_{\rm RSR}$), assuming $(U_{\odot},V_{\odot}, W_{\odot})= (11.10, 12.24, 7.25)$\,\kms~from \cite{Schonrich} and $R_0=8.34$\,kpc from \cite{Reid}. For the circular velocity around the Galactic centre, $\Theta_{\text{0}}$ was adopted as 240\,\kms~from \cite{Reid}, and $\Theta_{\text{R}}$ was computed from the \textit{MW2014} Galactic potential \citep{Bovy2015}. This is axisymmetric and composed of a spherical bulge, a Miyamoto-Nagai disc, and a halo with a Navarro-French-White (NFW) profile \citep[][]{nfw97}.

The obtained radial velocities were combined with the mean proper motions from \gaia EDR3 and distances from \citet{cantatgaudin2020} to derive full spatial velocities with respect to the Galactocentric standard of rest $(V_\text{R}, V_{\phi}, V_\text{z})$ and to the regional standard of rest $(U_\text{s},V_\text{s},W_\text{s})$, being $U_\text{s}=-V_\text{R}$, $V_\text{s}=V_\phi-\Theta_R$ and
$W_\text{s}=V_\text{z}$. The results are listed in Table~\ref{tab:GSR_RSR}. For those clusters without $\sigma_{v_{\rm rad}}$ values, we assumed $\sigma_{v_{\rm rad}}$=1\kms: ASCC\,23, Alessi\,44, NGC\,581, NGC\,1027, NGC\,1750, NGC\,2186, and Stock\,1. The uncertainties were estimated with a Monte Carlo experiment with 10$^5$ realisations, taking the uncertainties in radial velocities and distances of the clusters into account. The clusters show $v_\text{RSR}$ values typical of the thin disc kinematics (i.e. from $-25$ to $+20$\,\kms), and therefore they are typical of the thin disc kinematics \citep{binneytremaine2008}.

\begin{table*}
\setlength{\tabcolsep}{1.25mm}
\begin{center}
        \caption{Values of $v_\text{GSR}$, $v_\text{RSR}$, $(U_\text{s}$, $V_\text{s}$, $W_\text{s})$, and $V_{\phi}$ determined for each cluster.}
        \label{tab:GSR_RSR}
        \begin{tabular}{lcccccc} 
                \hline
Cluster &  $v_\text{GSR}$  &   $v_\text{RSR}$ &   $U_\text{s}$      &   $V_\text{s}$ & $W_\text{s}$  & $V_{\phi}$\\
 & [\kms] & [\kms] & [\kms] & [\kms] &[\kms] & [\kms]\\
\hline
  ASCC\,23     & 31.0$\pm$1.0  & -15.9$\pm$1.0 & 19.9$\pm$1.0 & 8.3$\pm$0.5 & 5.9$\pm$0.6 & 246.8$\pm$0.6\\
  Alessi\,44   & 157.4$\pm$1.0 & -5.4$\pm$1.4  & -6.3$\pm$1.4 & -0.6$\pm$1.0 & 4.0$\pm$1.1 & 241.0$\pm$1.0\\
  Alessi\,62   & 219.3$\pm$0.1 & 20.5$\pm$0.9  & 4.3$\pm$1.5  & 20.6$\pm$0.4 & 7.0$\pm$0.6 & 261.8$\pm$0.4\\
  COIN-Gaia\,2 & 167.2$\pm$0.2 & -12.5$\pm$1.6 & 24.5$\pm$0.6 & 4.4$\pm$1.9  & -1.3$\pm$1.3 & 242.4$\pm$1.7\\
  COIN-Gaia\,6 & 140.5$\pm$0.3 & 4.0$\pm$3.5   & 7.6$\pm$2.3  & 17.0$\pm$3.5 & -6.4$\pm$2.0 & 250.1$\pm$2.8\\
  FSR\,0951    & -38.9$\pm$0.5 & 16.9$\pm$1.1 & -18.7$\pm$1.1 & 5.5$\pm$1.1 & 9.2$\pm$0.8 & 240.6$\pm$1.0\\
  King\,23     & -106.2$\pm$0.1 & 0.2$\pm$2.8 & 9.7$\pm$2.9   & -18.4$\pm$0.9 & 0.9$\pm$1.6 & 214.1$\pm$1.3\\
  NGC\,1027    & 124.9$\pm$1.0 & -25.7$\pm$1.8 & 31.4$\pm$1.3 & -2.6$\pm$1.7 & 11.9$\pm$1.2 & 235.1$\pm$1.5\\
  NGC\,1647    & -20.6$\pm$0.1 & -19.3$\pm$0.1 & 18.9$\pm$0.3 & 12.2$\pm$0.7 & 3.8$\pm$0.8 & 250.6$\pm$0.7\\
  NGC\,1750    & -18.5$\pm$1.0 & -21.8$\pm$1.0 & 22.1$\pm$1.0 & 9.3$\pm$0.7  & 1.5$\pm$1.0 & 247.4$\pm$0.8\\
  NGC\,2186    & -90.3$\pm$1.0 & -15.9$\pm$1.9 & 20.5$\pm$2.4 & -9.8$\pm$1.8 & -1.4$\pm$1.2 & 224.3$\pm$2.3\\
  NGC\,2281    & 32.3$\pm$0.1  & 13.3$\pm$0.2  & -14.6$\pm$0.7 & -1.6$\pm$1.6 & -1.6$\pm$1.5 & 237.1$\pm$1.7\\
  NGC\,2345    & -132.6$\pm$0.2 & 4.2$\pm$3.0  & -2.6$\pm$2.1 & -3.5$\pm$2.6 & -2.3$\pm$1.4 & 230.8$\pm$2.1\\
  NGC\,2358    & -175.0$\pm$0.9 & -2.2$\pm$1.5 & 8.2$\pm$1.5  & -4.9$\pm$0.8 & 2.1$\pm$0.5 & 233.6$\pm$0.7\\
  NGC\,581     & 144.7$\pm$1.0 & -8.2$\pm$3.2 & 7.0$\pm$2.7  & -4.6$\pm$1.9  & -1.2$\pm$1.2 & 230.5$\pm$1.5\\
  NGC\,7654    & 196.1$\pm$0.1 & -4.5$\pm$2.1 & 0.3$\pm$2.2  & -5.2$\pm$1.3  & 2.9$\pm$1.3 & 232.7$\pm$1.1\\
  Stock\,1     & 205.1$\pm$1.0 & -9.0$\pm$1.1 & -14.6$\pm$1.8 & -2.5$\pm$1.0 & -3.6$\pm$1.2 & 238.3$\pm$1.0\\
  Trumpler\,2  & 157.9$\pm$0.1 & 6.5$\pm$1.0  & -6.6$\pm$1.9 & 1.5$\pm$1.0  & -7.2$\pm$1.5 & 240.1$\pm$1.1\\
  UBC\,54      & 77.7$\pm$0.1  & -8.8$\pm$1.0 & 3.3$\pm$1.9 & -14.9$\pm$2.3 & 5.7$\pm$1.3 & 222.3$\pm$2.6\\
\hline
\end{tabular}
\end{center}
\end{table*}

\begin{figure}
\centering
\includegraphics[width=\columnwidth]{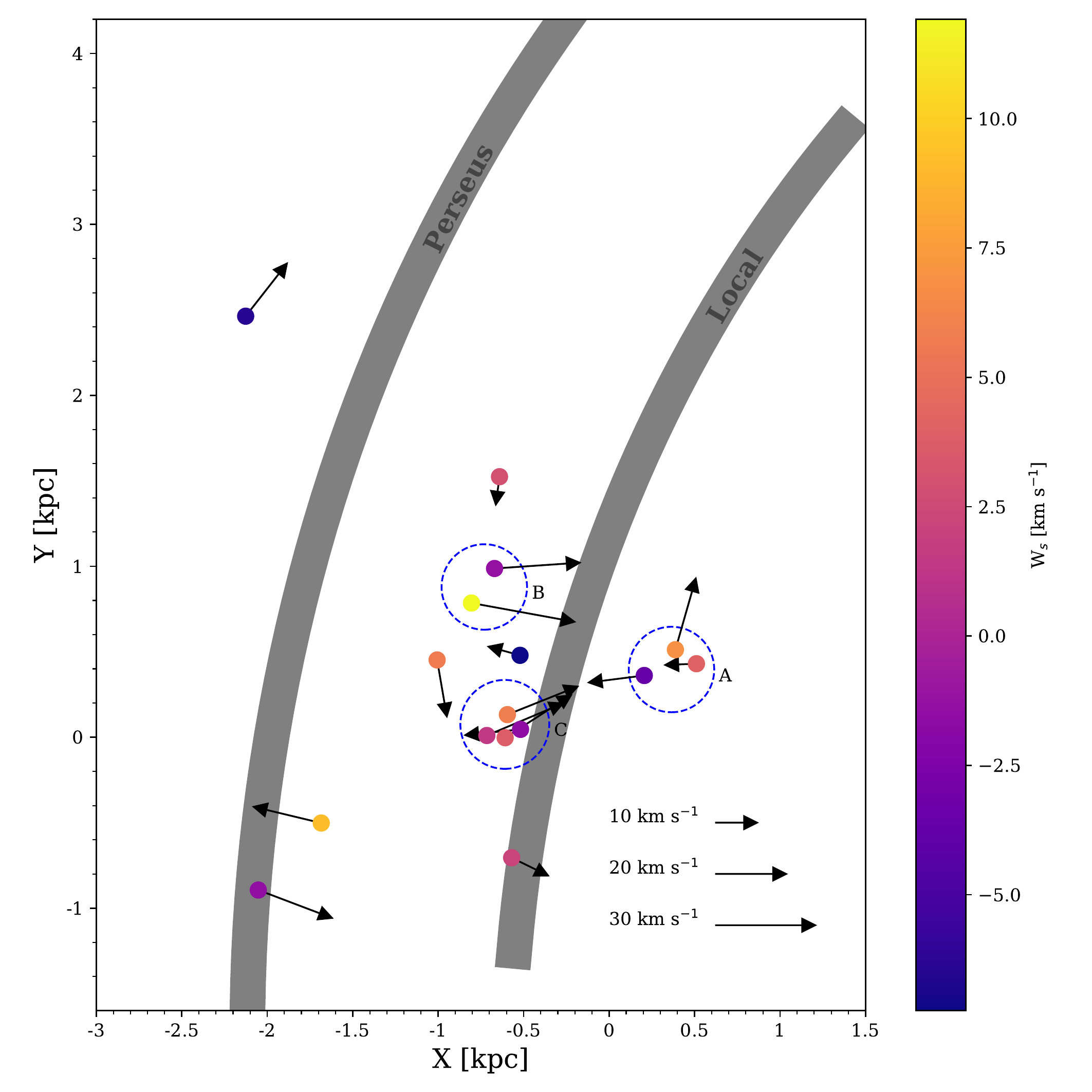}
\caption{Projection on the Galactic plane of the position and velocity with respect to the regional standard of rest of the clusters in our sample. The arrow sizes are proportional to $v_\text{RSR}$ as shown. The points are coloured as a function of $W_{\rm s}$. The shadow grey areas represent the spiral arms modelled by \citet{Reid}. }
\label{fig:ux_vy}
\end{figure}

Figure~\ref{fig:ux_vy} shows the projection on the Galactic plane of the position and velocity with respect to the regional standard of rest of the clusters in our sample. Although the projection of the velocity of Alessi\,44 and Stock\,1 are similar (group A in Fig.~\ref{fig:ux_vy}), their vertical components, $W_{\rm s}$, are not in agreement. The same happens for the pair COIN-Gaia\,2 and NGC\,1027 (group B). The trio ASCC\,23, NGC\,1647, and NGC\,1750 (group C) share similar components for $U_{\rm s}$, $V_{\rm s}$ and $W_{\rm s}$, and their ages range from 0.23 to 0.39\,Ga.

\subsection{Open clusters' orbits}

\begin{figure}
\centering
\includegraphics[width=\columnwidth]{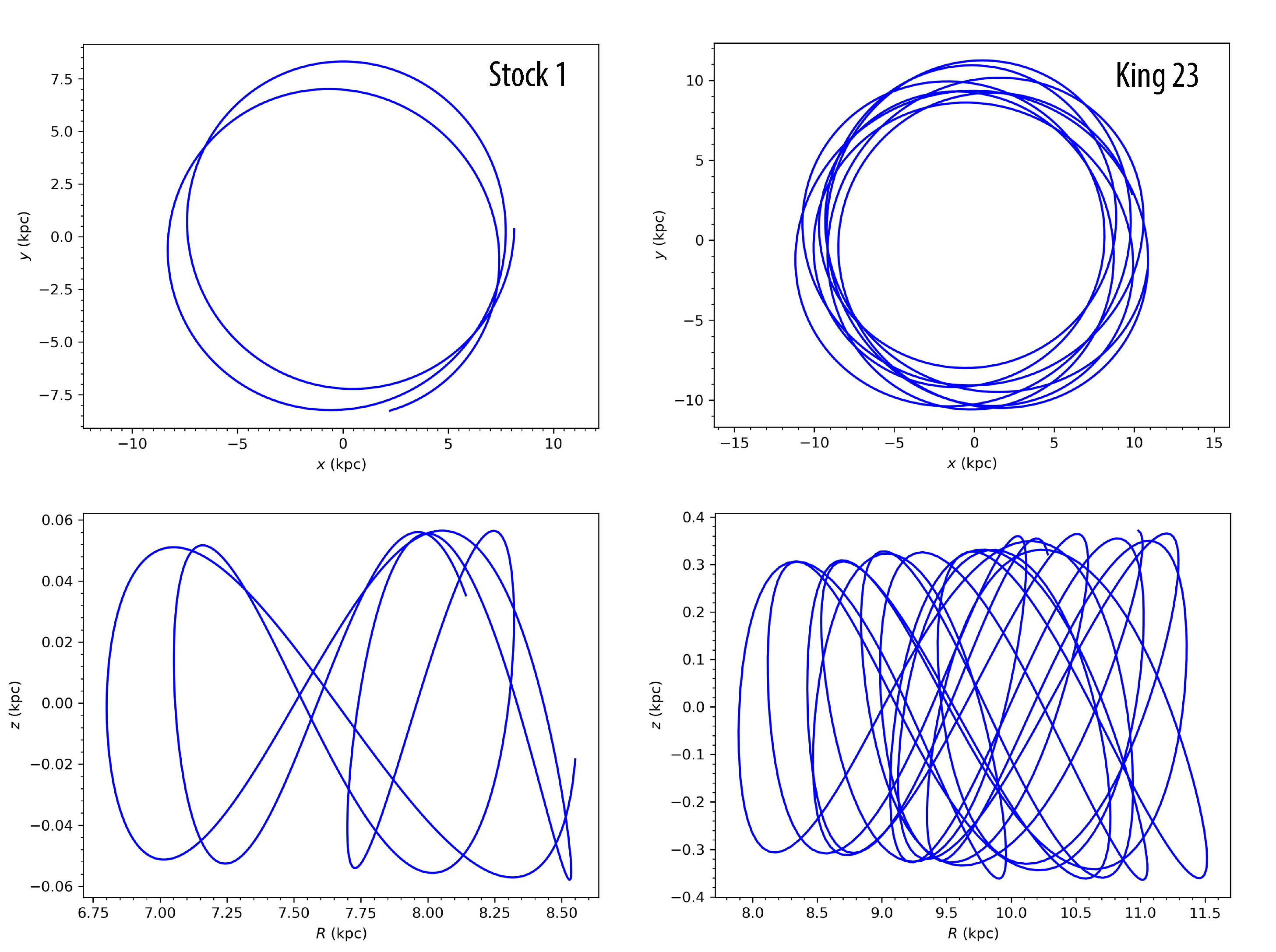}
\caption{Example of orbits for Stock\,1 (left) and King\,23 (right) with an age of 420\,Ma and 1.95\,Ga, respectively.}
\label{fig:ostta_orbit}
\end{figure}

The orbits of the OCs in our sample were integrated using the python \textit{galpy} package \citep{Bovy2015} using the same \textit{MW2014} Galactic potential described above, but we added two non-axisymmetric components: a bar and spiral arms. The bar was characterised with a Ferrers potential \citep{Ferrers1877} with $n=2$; the semi-major, middle, and minor axes are fixed to 3\,kpc, 0.35\,kpc, and 0.2375\,kpc, respectively. The bar mass is $10^{10}$\,M$_{\odot}$ \citep{Romero-Gomez2015} and a constant pattern speed was fixed to $\Omega$=42\,\kms\,kpc$^{-1}$ \citep{Bovy2019}, which puts co-rotation at $R=5.6$\,kpc and the outer Lindblad resonance at $R=9$\,kpc. The angular orientation of the bar with respect to the Sun-Galactic centre line is $20^{\circ}$ \citep[][and references therein]{Romero-Gomez2011}. For the spiral arms, we took the potential from \citet{Cox-Gomez2002}, assuming two arms with an amplitude of $0.4$ and a pattern speed of $\Omega=21$\,km\,s$^{-1}$\,kpc$^{-1}$ \citep[e.g.][]{Antoja2011}, which puts co-rotation at $R=10.6$\,kpc. 

We integrated the orbit backwards in time during the age of the cluster with a step of 2\,Ma. The uncertainties of the derived orbital parameters were estimated using Monte Carlo sampling, assuming Gaussian distributions for radial velocities, proper motions, distances, and their respective uncertainties. The derived orbital parameters and their uncertainties are listed in Table~\ref{tab:orb_par}. Figure~\ref{fig:ostta_orbit} shows an example of the orbits obtained for Stock\,1 (left) and King\,23 (right).

\begin{figure}
\centering
\includegraphics[width=\columnwidth]{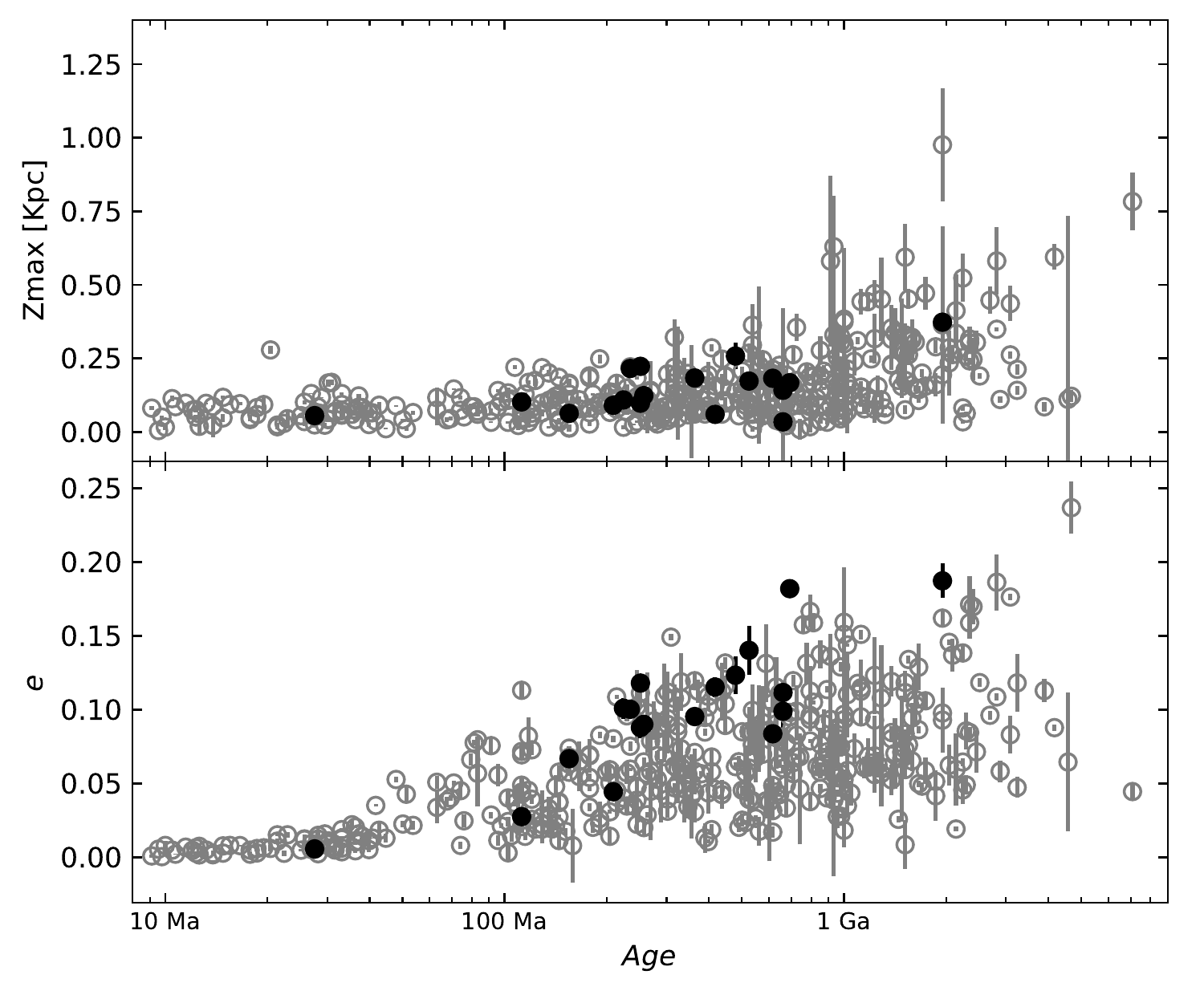}
\caption{Run of $z_{max}$ (top) and eccentricity (bottom) as a function of age for the OCs from the high-quality sample (grey) of \citet{tarricq2021} and for the OSTTA clusters studied here (black).}
\label{fig:zmax_e-age}
\end{figure}

It is far from the scope of this paper to perform a detailed analysis of the derived orbits due to the fact that the average radial velocities of several clusters have to be confirmed with the observations of more members. However, it is interesting to compare the derived orbital properties to the majority of the OC sample. To do that, we used the values derived by \citet{tarricq2021} who integrated the orbits using a similar procedure as the one used here for a sample of more than 1300 clusters.

In the top panel of Fig.~\ref{fig:zmax_e-age}, we plotted the run of maximum height above the plane $z_{\rm max}$ as a function of age for our clusters overplotted to the \citet{tarricq2021} high-quality sample. One can see the signs of vertical heating \citep{spitzer1951,jenkins1990} and that our clusters follow the general trend.

In the case of eccentricity, $e$ (bottom panel of Fig.~\ref{fig:zmax_e-age}), the studied clusters tend to be among the systems with the largest eccentricities for their given ages, and even above them, as is the case of the oldest ones: Alessi\,62 and King\,23. However, this may be explained by the fact that the orbits of the comparison sample by \citet{tarricq2021} were computed only with \textit{MW2014} potential without adding non-axisymmetric components. In fact, \citet{tarricq2021} suggest that the OCs may be born in nearly circular orbits, but during their lives, the eccentricity of their orbits increases due to the perturbations with the non-axisymmetric Galaxy components, such as with spiral arms.

\subsection{Open clusters' abundance ratios}

As in the case of kinematics, it is interesting to check if the derived chemical abundances follow the trends described by the majority of open clusters. For this comparison, we used the large homogeneously analysed sample obtained by \citet{spina2021_galahOC} from the APOGEE and GALAH Galactic surveys. This sample contains 134 clusters for which abundances for 21 elements, from C to Eu, are provided. Unfortunately, this sample does not provide abundances for two of the elements studied, Sc and Ce. For this reason, we also used the abundances derived by \citet{casamiquela21abun} from spectra acquired with different instruments for stars around the red clump region belonging to 47 systems. Although this is a more limited sample with clusters in a radius of 500\,pc with ages older than 200\,Ma, it is still useful for our purpose. Finally, in the case of Ce, we added the comparison with the results obtained by \citet{salessilva+2022} from APOGEE infrared spectra.

\begin{figure*}
\centering
\includegraphics[width=\textwidth]{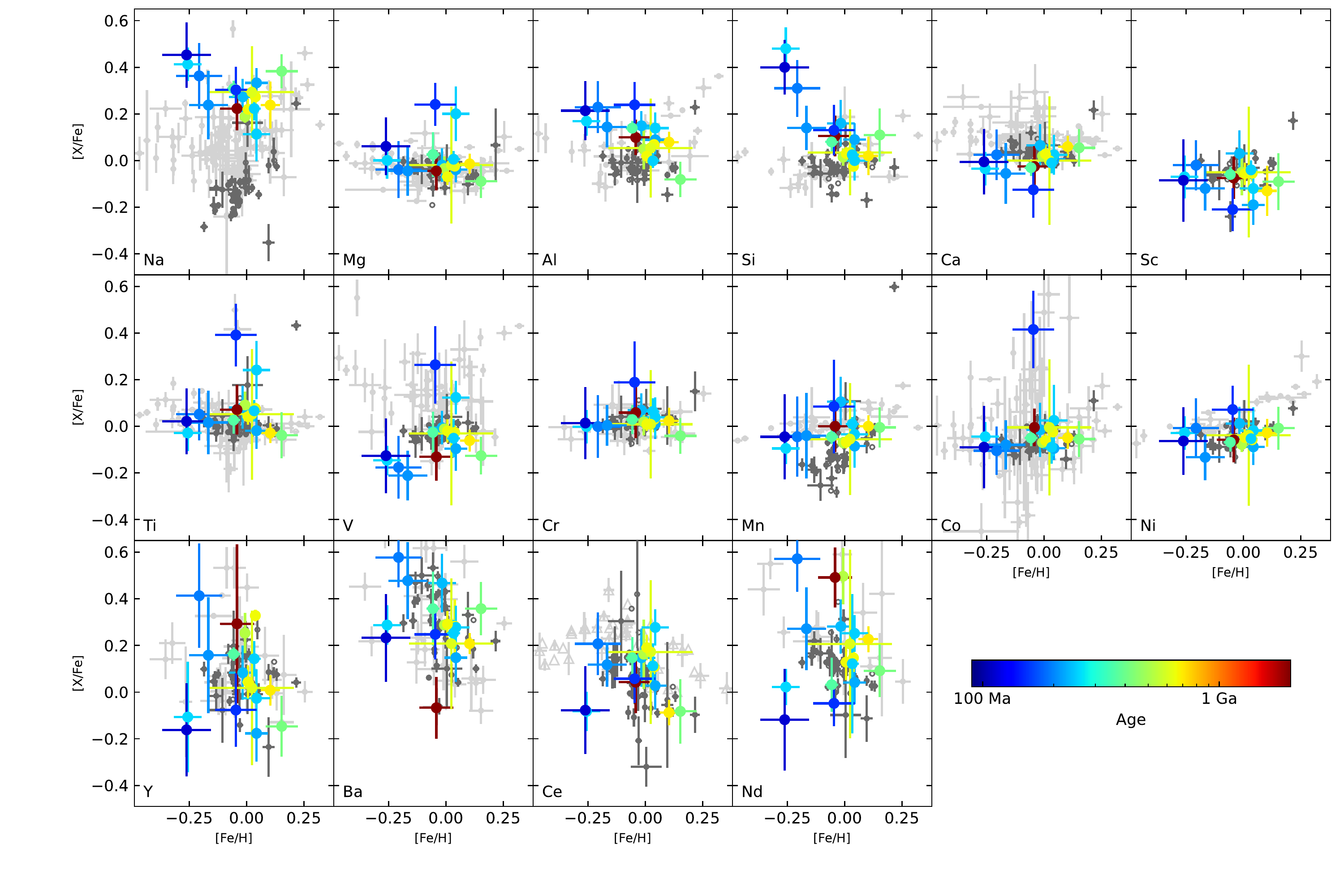}
\caption{Abundance ratios [X/Fe] versus [Fe/H] for the studied clusters, colour-coded as a function of their ages. Light and dark grey points are the OCs sampled by \citet{spina2021_galahOC} and \citet{casamiquela21abun}, respectively. In the case of Ce, light grey triangles are the values obtained by \citet{salessilva+2022}.}
\label{fig:ratios}
\end{figure*}

The different [X/Fe] ratios as a function of [Fe/H] for our sample are shown in Fig.~\ref{fig:ratios} and are colour-coded as a function of the age. The samples of \citet{spina2021_galahOC} and \citet{casamiquela21abun} are in light and dark grey, respectively.

The Na comparison is intriguing. There is no agreement between the two comparison samples. For the 14 clusters in common to both studies, the \citet{spina2021_galahOC} Na abundances are on average 0.24\,dex higher than the  \citet{casamiquela21abun} ones. Moreover, the Na abundances derived for the clusters studied here are higher than the \citet{spina2021_galahOC} ones. One explanation could be that our analysis did not take into account non-LTE corrections for the derived abundances, as is the case for \citet{spina2021_galahOC}. However, in regards to Na, the abundances derived in LTE, as is also the case here, could be overestimated by up to 0.5\,dex according to \citet{Lind+2011}. Moreover, as discussed by \citet{Casamiquela+2020} \citep[see also][]{Smiljanic+2016,Smiljanic+2018}, the Na abundances could be affected by internal mixing in the surface of massive giants, which is related to the age of the clusters being more important for younger objects as observed in Fig.~\ref{fig:ratios}. The comparison for another light proton-capture element such as Al, which is less affected by non-LTE corrections in our case \citep[e.g.][]{nordlander2017}, is in better agreement.

It is surprising that the three most metal-poor clusters studied, NGC\,1027, NGC\,1750, and Trumpler\,2, are enhanced in Si by about 0.4\,dex, but not in the other $\alpha$-elements analysed (Mg, Ca, and Ti). The abundances for two of these clusters, NGC\,1027 and NGC\,1750, are based on only one star each, and therefore this result may be taken with caution. However, in the case of the young, 110\,Ma, Trumpler\,2 system, we have analysed two stars and both have similar Si-enhanced abundances. The most probable explanation for this is the existing bias in Fe abundances as a function of \teff\ and \logg, and thus with age, particularly affecting the youngest stars in our sample (Sect.~\ref{sec:chemabs}). Since this bias tends to underestimate the Fe abundance of cooler stars, but not the Si abundance, we have probably obtained an underestimation of the [Fe/H] abundances and a net enhancement in [Si/Fe].
An enhancement in Si has been reported in another young (300\,Ma, older than the three previously mentioned clusters) yet more metal-rich ([Fe/H]=0.1\,dex) cluster,  NGC\,6705 \citep{Magrini+2014,Casamiquela+2018}, but not by other works based on APOGEE data \citep[e.g.][]{Donor+2020,spina2021_galahOC}. However, in this case, the Si enhancement is accompanied by an enhancement in Mg and O. The fact that no other [$\alpha$/Fe] enhancement is obtained in these three clusters makes us think that they are most probably not $\alpha$-enhanced objects. 

Regardless, the existence of field stars with [Fe/H] in the range of the OCs and enhancements in $\alpha$-element abundances have also been reported in the literature \citep[e.g.][]{Adibekyan+2011,Martig+2015,Chiappini+2015}. Three scenarios have been proposed for the origin of the young [$\alpha$/Fe]-enhanced stars. The first assumes that the younger ages, which are determined from their masses, are wrong and that they are in fact older since their masses are higher than expected due to mass transfer from a binary companion, for instance, or they resulted from the merger of two stars \citep[e.g.][]{Martig+2015,izzard2018}. This is not the case of OCs where ages are more accurately determined from their colour-magnitude diagrams. However, stars with a higher mass than expected have been found in at least two OCs where masses could be determined from asteroseismology. Those high masses were explained as results of a merger or mass transfer in binary systems; for more information, readers can refer to \citet{handberg+2017} for NGC\,6819 and \citet{brogaard+2021} for NGC\,6791. The second scenario suggests that these objects are genuinely young, that they formed near the region of co-rotation of the Galactic bar, and that they have migrated to the solar neighbourhood \citep[][]{Chiappini+2015}. However, the orbits determined in previous sections suggest that the clusters were probably born roughly in the inner radius, but far enough away from the bar. An alternative is local self-enrichment due to the explosion of a supernova type II in a giant molecular cloud, as proposed by \citet{Magrini+2015}. According to these authors, a single explosion of a massive star with a mass in the range between 18 and 25\,M$_{\odot}$ should be able to explain the Si abundances observed in these clusters. Moreover, the yields reported by \citet{woosley1995} supported the higher enrichment in Si with respect to the other $\alpha$-elements. 

In the case of Ca, the abundances obtained here are slightly lower than those obtained by \citet{spina2021_galahOC}, suggesting the existence of a zero-point between both samples. It is expected that [Ca/Fe]$\sim$0.0\,dex at solar metallicity \citep[e.g.][]{Magrini+2014,carrera2019apo}. Therefore, the values reported by \citet{spina2021_galahOC} seem slightly overestimated.

A similar behaviour is observed in the case of V. For this element, it is interesting that the two metal-poor clusters in our sample have slightly lower abundances, although with large uncertainties. For the other refractory element, Sc, there is good  agreement with the values reported by \citet{casamiquela21abun}. Unfortunately, \citet{spina2021_galahOC} do not provide Sc abundances to compare with the values obtained here.

For the Fe-peak elements studied (Cr, Mn, Co, and Ni), there is agreement with the comparison samples within the uncertainties. The only noticeable feature is the enhancement of $\sim$0.4\,dex in Co for NGC\,7654. This cluster also shows a similar enhancement in Ti. A single giant star was observed in this cluster, and therefore these results should be taken with caution. Moreover, there are doubts about the true membership of this star because the isochrones are not able to reproduce its positions in either the colour-magnitude (Fig.~\ref{fig:dcms}) or Kiel (Fig.~\ref{fig:kiel}) diagrams, as discussed before. The \citet{spina2021_galahOC} sample has other clusters with similar enhancements in Ti and Co. However, those systems with high Co abundances are not the same with a high Ti content.

Finally, the abundances derived for the four neutron capture elements analysed (Y, Ba, Ce, and Nd) are in agreement with the \citet{spina2021_galahOC} and \citet{casamiquela21abun} comparison samples taking into account the largest uncertainties in comparison to other elements involved in their determinations. In fact, the abundances of Ce and Nd are based on the measurements of a single weak line in each case: 527.4229 and 531.981\,nm for Ce and Nd, respectively. The only noticeable feature is that two of the more metal-poor clusters (NGC\,1750 and Trumpler\,2) seem slightly out of the global trend. This result has to be taken with caution due to the large error bars, which prevent us from discussing this further. In the case of Ce, there is a clear shift between both our sample and the \citet{casamiquela21abun} samples with the values obtained by \citet{salessilva+2022} from APOGEE near-infrared spectra, which are slightly larger at any [Fe/H] value. \citet{salessilva+2022} quantified the difference with \citet{casamiquela21abun} in 0.16\,dex. The differences may be due to the different solar reference abundances used in each case. Regardless, the trend reported by \citet{salessilva+2022} of increasing [Ce/Fe] with decreasing [Fe/H] until about [Fe/H]$\sim$0.2\,dex is also the case for the results obtained here.

\subsection{Open clusters radial gradient}

Another interesting feature of OCs is the run of [Fe/H] as a function of Galactocentric distance, $R_{gc}$. It is widely accepted that [Fe/H] decreases as we move outwards, which seems to flatten when reaching a certain distance \citep[e.g.][]{Carrera+2011,Magrini+2017,donor2020dr16}. Unfortunately, the clusters studied here are located inside the breaking radius, and therefore they are not useful for investigating its location. However, it is interesting to compare the results obtained here with the majority of OCs. 

\begin{figure}
\centering
\includegraphics[width=\columnwidth]{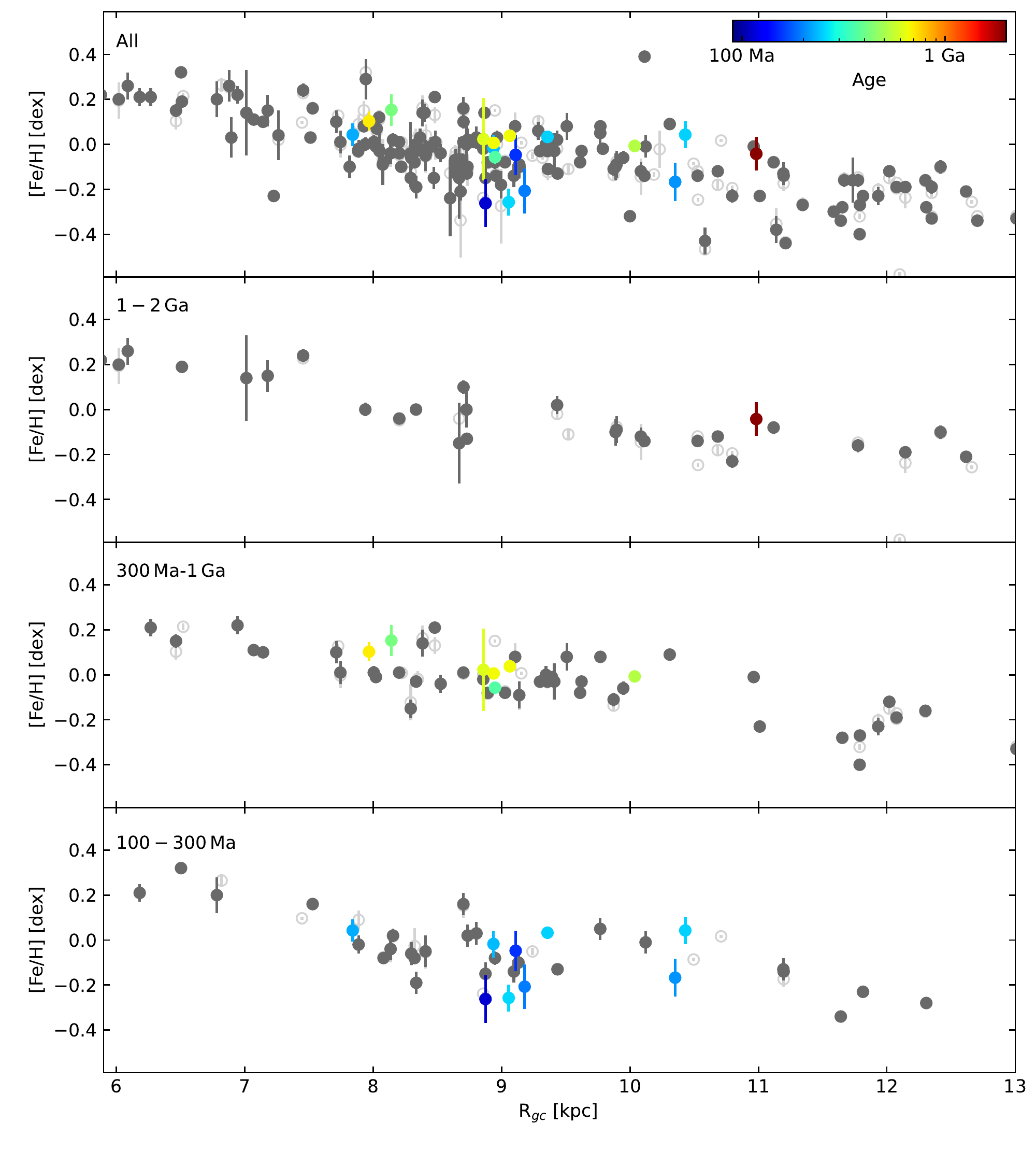}
\caption{Run of [Fe/H] versus $R_{gc}$ for the studied clusters, colour-coded as a function of their age. Dark and light grey points are the clusters in \citet{zhang_spa} and \citet{spina2021_galahOC}, respectively, which are plotted for comparison. The top panel shows the whole sample, while the remaining panels show three different age ranges.}
\label{fig:rad_gradient}
\end{figure}

In the top panel of Fig.~\ref{fig:rad_gradient}, we plotted the run of [Fe/H] versus $R_{gc}$ for the clusters studied here, which are colour-coded as a function of their age. As comparison samples, we have used the recent compilation obtained by \citet{zhang_spa}, which includes 157 clusters from different sources, and the sample obtained by \citet{spina2021_galahOC} from the APOGEE and GALAH surveys described above. They are the dark and light grey points in Fig.~\ref{fig:rad_gradient}. In general, the clusters studied here follow the trend described by the majority of open clusters. 
However, the existence of particular cases can be blurred when comparing all the clusters together, independently of their ages. In fact, it is known that the slope of the gradient changes with age, being steeper for the oldest clusters \citep[e.g.][]{Friel+2002,Carrera+2011,donati2015tr5,donor2020dr16,zhang_spa}. For this reason, in the rest of the panels of Fig.~\ref{fig:rad_gradient}, we plotted three different age ranges. The clusters studied here follow, within the uncertainties, the trends described by coeval systems without noticeable differences. 

\section{Conclusions}\label{sec:conclusion}

The OSTTA project was designed to provide high resolution follow-up spectroscopy of poorly studied open clusters. In this paper, we derived radial velocities for 41 stars belonging to 20 open clusters, including the following four systems recently discovered from \gaia data releases: COIN-Gaia\,2, COIN-Gaia\,6, Gulliver\,37, and UBC\,54. 

To our knowledge, our analysis provides the first radial velocity determination based on high resolution spectroscopy for most of the studied clusters. At the time of writing, only \gaia radial velocities were available for most of them. The radial velocities allowed us to detect four potential spectroscopic binaries and to revise the membership of each cluster, discarding four objects. After this procedure, we cannot be sure that we sampled a true member of Gulliver\,37 since the star observed in this cluster was reported as a variable in the literature.

Atmospheric parameters were determined for the 32 stars considered as cluster members from their radial velocities. Two stars have very low gravities, which is an indication of being supergiant stars. For two objects, we were not able to properly constrain the atmospheric parameters due to the low S/N of their spectra. Likewise, we were not able to constrain the atmospheric parameters of the star observed in NGC\,581, which was reported in the literature as an M supergiant. Abundances for 17 chemical species were determined in the remaining 28 stars belonging to 17 open clusters. To our knowledge, this is the first chemical abundance determination for most of them.

We investigated the behaviour of the studied clusters in the framework of the majority of the open clusters. All the studied clusters show typical thin disc kinematics, and their orbits are also within the ranges described by the majority of open clusters. The derived abundance ratios follow the general trends described by the majority of clusters. However, we found that three young clusters, NGC\,1027, NGC\,1750, and Trumpler\,2, are enhanced in Si without signs of enrichment in the other $\alpha$-elements studied (Mg, Ca, and Ti). This behaviour was only observed previously in another young cluster, NGC\,6705. There is no clear explanation for the abundances measured in these clusters. Finally, we confirmed that the [Fe/H] abundances derived for the  studied clusters follow the radial gradient traced by other clusters of similar ages.

\begin{acknowledgements}
      Based on observations made with the Nordic Optical Telescope, owned in collaboration by the University of Turku and Aarhus University, and operated jointly by Aarhus University, the University of Turku and the University of Oslo, representing Denmark, Finland and Norway, the University of Iceland and Stockholm University at the Observatorio del Roque de los Muchachos, La Palma, Spain, of the Instituto de Astrof\'{\i}sica de Canarias. We acknowledge funding from the Italian MIUR though Premiale 2016 MITiC.
      This work was (partially) funded by the Spanish MICIN/AEI/10.13039/501100011033 and by "ERDF A way of making Europe" by the “European Union” through grant RTI2018-095076-B-C21, and the Institute of Cosmos Sciences University of Barcelona (ICCUB, Unidad de Excelencia ’Mar\'{\i}a de Maeztu’) through grant CEX2019-000918-M.
      This work presents results from the European Space Agency (ESA) space mission \gaia. \gaia data are being processed by the Gaia Data Processing and Analysis Consortium (DPAC). Funding for the DPAC is provided by national institutions, in particular the institutions participating in the Gaia MultiLateral Agreement (MLA). The Gaia mission website is https://www.cosmos.esa.int/gaia. The Gaia archive website is https://archives.esac.esa.int/gaia.
      This research made use of Astropy,\footnote{http://www.astropy.org}  \citep{astropy:2013, astropy:2018}, Matplotlib \citep[][]{matplotlib}, Sk-learn \citep{scikit-learn} python packages. This research has made use of the SIMBAD database operated at CDS, Strasbourg, France \citep{simbad}, and TOPCAT \citep{topcat2005}.
\end{acknowledgements}

%
%

\bibliographystyle{aa} 
\bibliography{rcarrera,biblio2_v2}

\appendix

\section{Notes on individual clusters}\label{apex:indvnotes}

All the \gaia identifications used in this section refer to \gaia EDR3, except where otherwise specified.

\paragraph{ASCC\,23} The radial velocity of the star observed in this cluster, with $p$=1, is in good agreement within the uncertainties with the values reported by \citet{soubiran2018gaiadr2_rv}  and \citet{tarricq2021} derived from six and nine stars, respectively. Therefore, we consider it as representative of the radial velocity of the cluster.

\paragraph{Alessi\,44} A single exposure has been acquired for the star observed in Alessi\,44. The derived radial velocity is in agreement with the values available in the literature by \citet{soubiran2018gaiadr2_rv}  and \citet{tarricq2021} derived from seven and nine stars, respectively. Therefore, although this star has $p$=0.5, we consider it as representative of the cluster.

\paragraph{Alessi\,62} Two out of three stars observed in Alessi\,62 have $p$=1 and compatible radial velocities. The third star, 45193335743607072, was observed; although, it has a negligible membership probability. Its derived radial velocity is very different from the other two and from the value provided by \gaia DR2 for it. Coupled with the large radial velocity uncertainty of $\sim$5\kms~reported by \gaia, this suggests that this object might be a spectroscopic binary. The average velocity derived from two member stars is in good agreement with the values found by \citet{soubiran2018gaiadr2_rv}  and \citet{tarricq2021} from 12 and ten stars, respectively.

\paragraph{COIN-Gaia\,2} Two of the observed stars in this cluster have $p$=1. The other one, 423486180567313024, has a very low value, $p$=0.05. In fact, its radial velocity differs by $\sim$3\kms~from that of the other two. We consider this star as a non-member, but it could also be a spectroscopic binary.

\paragraph{COIN-Gaia\,6} The three observed stars have very high astrometric membership probabilities and similar radial velocities among them. However, the average radial velocity derived for this cluster significantly differs from the value reported by \citet{tarricq2021}: -47.26$\pm$2.41\kms, derived from \gaia DR2 measures of three stars. There are two stars in common between both samples, 506224812820610176 and 506223988186933888. Whereas there is good agreement between the radial velocities derived here and those provided by \citet{tarricq2021} for one star, there is a difference of $\sim$3\kms~for the other. The \gaia DR2 radial velocity uncertainty for this object is 2.69\kms. This, together with the small discrepancy between both studies, could be a hint that this star is a spectroscopic binary. The additional star used by \citet{tarricq2021}, which is not in common with our sample, has a significantly lower radial velocity $-51.3\pm0.2$\kms~in \gaia DR2 and a lower astrometric membership probability, 0.47. This slightly discrepant star as well as the previously mentioned discrepant object in common are the cause of the difference in the average $v_{rad}$. In our study, however, the three radial velocities are in very good agreement and with consistent low uncertainties, thus we consider the three stars to be real cluster members.   

\paragraph{FSR\,0951} The three stars observed have 0.5$\leq$p$\leq$0.9. The star 3368695854168200192, with the lower astrometric membership probability, $p$=0.5, has a slightly different $v_{rad}$. The \gaia DR2 radial velocity for this star, 46.15$\pm$0.38\kms, is in good agreement with our measurement. Comparing this with \gaia EDR3 mean proper motions and parallaxes, this star has $\mu_{\alpha}$ outside three sigma from the mean cluster value listed in Table~\ref{tab:ocs}. We discarded this object in our analysis. From the remaining stars, we obtained an average radial velocity of 43.80$\pm$0.48\kms, which is compatible with the value reported by \citet{soubiran2018gaiadr2_rv} and \citet{tarricq2021} from five stars in both cases.  

\paragraph{Gulliver\,37} The only star sampled in this cluster, 2024469226291472000, has $p$=0.9. The radial velocity obtained here, 21.92$\pm$0.02\kms, is quite different from the value provided by \gaia DR2 of  2.5$\pm$8.2\kms~for the same star. Recently, this star was observed in the framework of the Stellar Population Astrophysics (SPA) project by \citet{zhang_spa} who obtained a radial velocity of \minus4.59$\pm$0.17\kms. Due to the large velocity dispersion reported by \gaia DR2 and the discrepancy among the different radial velocity determinations available, we consider this object as a potential spectroscopic binary, although our spectrum did not show either significant wide lines or double lines. Therefore, we cannot ensure that it is a real cluster member, and we excluded this cluster from our analysis.

\paragraph{King\,23} The three stars observed in King\,23 have $p\geq$0.8. The derived radial velocity for one of them, 3109989396744298624, differs by $\sim$3\kms~from the other two. A similar difference between the radial velocities of the discrepant star and the other two is observed in the values provided by \gaia DR2. Since there is no hint of it being a spectroscopic binary, we discarded it from our analysis. The average radial velocity provided by \citet{soubiran2018gaiadr2_rv} and \citet{tarricq2021} for this cluster, obtained from the \gaia DR2 velocities of four stars, slightly differs from the value found here because of the contamination of the discrepant star. 

\paragraph{NGC\,581} The single star observed has $p$=0.8. This star was catalogued as an M supergiant (M0.5Ib-II) by \citet{keenan1989} and reported as variable by \citet{Mermilliod+2008}. The derived radial velocity is compatible with the \gaia DR2 value. Neither our measure nor the \gaia DR2 one shows a sign of radial velocity variability. The average values for this cluster provided by \citet{soubiran2018gaiadr2_rv}, for this same star, and by \citet{tarricq2021}, for two stars, are in agreement with the value found here. We assumed that this star is a real cluster member, although this should be confirmed in the future by sampling other stars. However, we discarded this object in the chemical analysis because it is an M supergiant.

\paragraph{NGC\,1027} The two stars observed in this cluster have $p$=1, 465682692367536384, and 0.8, 465853425905007872, respectively. However, we found very different radial velocities for each of them. \gaia DR2 also provides radial velocities for these stars which are in good agreement with our individual measurements within the uncertainties. There is a significant difference between the values provided for this cluster by \citet{soubiran2018gaiadr2_rv} from one star -4.06$\pm$0.31\kms, and \citet{tarricq2021} from six stars -36.57$\pm$10.83\kms. The value found by \citet{tarricq2021} is in agreement with the radial velocity derived for the star 465682692367536384 within the uncertainties. For this reason, we consider this star as a real member of the cluster, although this assumption should be confirmed in the future. 

\paragraph{NGC\,1647} The two stars observed in this cluster have $p$=1 and compatible radial velocities within the uncertainties. The derived radial velocities are in agreement with the values reported by \gaia DR2 and \citet{Mermilliod+2008} within the uncertainties. The obtained average radial velocity is in good agreement with the values provided by \citet{soubiran2018gaiadr2_rv} and \citet{tarricq2021} from 16 and 21 stars, respectively.

\paragraph{NGC\,1750} The two observed stars have $p$=1 (3418663507987229184) and $p$=0.8 (3418709412597592576). However, their radial velocities differ by $\sim$3\kms. \gaia DR2 provided radial velocities for these two objects. There is good agreement for star 3418663507987229184. However, for 3418709412597592576, there is a difference of about 2\kms; although, the \gaia DR2 value has an uncertainty of 1.1\kms, which could be a sign that this star is a spectroscopic binary. The value that we found for the first star is also in agreement with the average radial velocity provided by \citet{soubiran2018gaiadr2_rv} for this cluster, -10.38$\pm$2.31\kms~, from seven stars, including the two observed here. \cite{tarricq2021} provide a significantly different value, -7.45\kms, but with a larger uncertainty, 6.64\kms, from 13 stars. The star 3418709412597592576 is slightly displaced from the isochrone, but we see that the isochrone is not a perfect match for the cluster sequence, mainly at the main sequence turn-off (see Fig.~\ref{fig:dcms}). Therefore, with all the information in hand, we assume that star 3418663507987229184 is representative of the cluster, although this assumption should be confirmed in following studies.

\paragraph{NGC\,2186} The single observed star has $p$=1. The obtained radial velocity is compatible within uncertainties with the value determined by \citet{Mermilliod+2008} and with that provided by \gaia DR2. Moreover, this is the only star used by \citet{soubiran2018gaiadr2_rv} as a representative of this cluster. The value used by \citet{tarricq2021} for this star is 20.98$\pm$0.78\kms, which is also compatible with our value. These authors determined an average radial velocity for this cluster of 21.71$\pm$4.01\kms~ obtained from three stars. All together, we consider the observed star as a real cluster member.

\paragraph{NGC\,2281} The two stars observed in this system have $p$=1 and similar radial velocities within the uncertainties. Moreover, the obtained velocities are in agreement with \gaia DR2, \citet{Mermilliod+2008}, and \citet{luck2014}, again, within the uncertainties. The obtained average radial velocity is in agreement with the values provided by \citet{soubiran2018gaiadr2_rv} and \citet{tarricq2021} from 42 and 40 stars, respectively, and also from the values obtained by \citet{smiljanic2018_AA616} for two objects in this cluster.

\paragraph{NGC\,2345} We have observed three objects in NGC\,2345 with $p\geq$0.9. One of them, 3044669506889464320, was reported as a spectroscopic binary by \citet{Mermilliod+2008}. The derived radial velocity for this star significantly differs from the values for the other two. A similar difference is observed in the \gaia DR2 radial velocities, which are in agreement with the values obtained here. The average radial velocity obtained by discarding this object, 58.41$\pm$0.15\kms, slightly differs from the values reported by \citet{soubiran2018gaiadr2_rv}, 60.94\kms, and \citet{tarricq2021}, 63.27\kms, from six stars in both cases, but they reported large dispersions of 4.89 and 2.43\kms, respectively. It is worth mentioning that \citet{alonso_santiago2019} report an average radial velocity for this cluster of 58.5$\pm$0.5\,\kms~from four stars, including the two used here.  

\paragraph{NGC\,2358} The three stars observed have 0.6$\leq p \leq$0.9. The radial velocities of the three objects are similar within the uncertainties, and also they are compatible with the average values derived from \gaia DR2 velocities. In fact, the average radial velocity obtained, 28.31$\pm$0.85\kms, is compatible within the uncertainties with the values provided by \citet{soubiran2018gaiadr2_rv}, 27.57$\pm$1.03\kms, and \citet{tarricq2021}, 27.57$\pm$0.6\kms, from the same three stars.

\paragraph{NGC\,7654} The two stars observed in this cluster have $p$=1 and similar radial velocities within the uncertainties between them and with the values provided by \gaia DR2. The average radial velocity is in agreement with the values provided by \citet{soubiran2018gaiadr2_rv} and \citet{tarricq2021} because they were obtained from the \gaia DR2 velocities for the same stars. Both stars are far away from the expected location for the red clump for the age of this cluster (Fig.~\ref{fig:dcms}), suggesting a younger age. Recently, \citet{akbulut2021} revised the age of this cluster to a slightly younger value of 120\,Ma, which is still insufficient to reproduce the position of these stars in the colour-magnitude diagram. The same authors highlight that in order to reproduce the position of these stars, an isochrone of about 40\,Ma would be needed. In this case, the observed stars would be supergiants. Owing to the agreement in the radial velocities of the two observed stars, and also in the derived chemical abundances (see next section), we consider them as real clusters members. However, this is an assumption that should be taken with caution until their membership is confirmed with the radial velocities of other objects in the main sequence. 

\paragraph{Stock\,1} We have acquired a single exposure for one star in this cluster, which has $p$=1. The derived radial velocity is in good agreement with the values determined from \gaia DR2 by \citet{soubiran2018gaiadr2_rv} and \citet{tarricq2021} from 30 and 29 objects, respectively. Therefore, we consider the star as a real cluster member and the derived radial velocity as representative of the cluster.

\paragraph{Trumpler\,2} The two stars observed in this system have $p$=0.9 and similar radial velocities within their uncertainties and also with the values provided by \gaia DR2. The derived average radial velocity is in good agreement with the values obtained by \citet{soubiran2018gaiadr2_rv} and \citet{tarricq2021} from  11 and 12 objects, respectively. One star in this cluster was sampled by \citet{smiljanic2018_AA616} who found a compatible radial velocity with the value obtained here.

\paragraph{UBC\,54} The two stars observed in this cluster have $p$=1. The obtained radial velocities are in agreement within the uncertainties and also with \gaia DR2. The average radial velocity is compatible within the uncertainties with the value provided by \citet{tarricq2021} from the same two stars measured by \gaia DR2.

\section{Some extra material}

\begin{table*}
\setlength{\tabcolsep}{0.25mm}
\caption{Observing log and radial velocities}
\begin{tabular}{lcccccccccccl}
   \hline
   Cluster & Gaia ID EDR3 & RA & DEC & G-band & S/N & N & texp & Date  & $v_{\rm rad}$ & $v_{\rm scatter}$ & $v_{\rm err}$ & Notes \\
    &  & deg & deg & mag & pix$^{-1}$ & & s & & km s$^{-1}$ & km s$^{-1}$ & km s$^{-1}$ & \\
   \hline
   ASCC 23 & 968512002808979456 & 95.039717 & 46.646874 & 7.747 & 91.4 & 3 & 120 & 14/12/2018 & -13.347 & 0.019 & 0.025 & \\
   Alessi 44 & 4239077200124649216 & 294.322473 & -0.492237 & 7.873 & 98.1 & 1 & 180 & 24/04/2019 & -9.578 &  & 1.108 & \\
   Alessi 62 & 4519333643079413760 & 284.069306 & 21.548663 & 9.926 & 137.9 & 3 & 1800 & 18/04/2019 & 12.873 & 0.009 & 0.013 & \\
   Alessi 62 & 4519335601584434048 & 284.050962 & 21.65614 & 9.687 & 143.6 & 3 & 1800 & 18/04/2019 & 13.054 & 0.014 & 0.049 & \\
   Alessi 62 & 4519300898247954560 & 283.905531 & 21.265434 & 9.911 & 155.5 & 3 & 1800 & 19/04/2019 & 21.1 & 0.009 & 0.025 & SB?/NM\\
   COIN Gaia2 & 423486184877336448 & 15.076727 & 55.375846 & 10.181 & 87.8 & 3 & 1900 & 12/12/2018 & -35.512 & 0.018 & 0.048 & NM\\
   COIN Gaia2 & 423487731065542272 & 14.998977 & 55.45639 & 10.69 & 108.8 & 3 & 2100 & 12/12/2018 & -32.721 & 0.02 & 0.039 & \\
   COIN Gaia2 & 423486184872160128 & 15.086556 & 55.367102 & 10.736 & 102.8 & 3 & 2100 & 12/12/2018 & -32.39 & 0.023 & 0.018 & \\
   COIN Gaia6 & 506223988186933888 & 28.052953 & 58.561446 & 10.377 & 85.6 & 3 & 1500 & 14/12/2018 & -42.319 & 0.002 & 0.009 & \\
   COIN Gaia6 & 506224812820610176 & 28.049522 & 58.642991 & 13.144 & 52.7 & 3 & 10800 & 14/12/2018 & -42.118 & 0.029 & 0.109 & SB?/low S/N\\
   COIN Gaia6 & 506214401820077056 & 28.156907 & 58.629604 & 13.331 & 48.1 & 3 & 10800 & 15/12/2018 & -42.811 & 0.004 & 0.158 & low S/N\\
   FSR 0951 & 3344673277448107776 & 95.603076 & 14.476332 & 11.526 & 132.7 & 6 & 9000 & 12,16/12/2018 & 43.462 & 0.042 & 0.025 & \\
   FSR 0951 & 3344692140944457088 & 95.351933 & 14.667711 & 11.574 & 133.8 & 6 & 9000 & 12,15/12/2018 & 44.137 & 0.042 & 0.034 & \\
   FSR 0951 & 3368695854168200192 & 95.689046 & 14.61152 & 11.352 & 136.7 & 6 & 8400 & 13,14/12/2018 & 46.279 & 0.027 & 0.032 & NM\\
   Gulliver 37 & 2024469226291472000 & 292.076836 & 25.381498 & 10.592 & 101.7 & 3 & 2700 & 19/04/2019 & 21.92 & 0.024 & 0.051 & SB\\
   King 23 & 3109989396744298624 & 110.428062 & -0.963394 & 12.123 & 91.9 & 3 & 8100 & 13/12/2018 & 56.471 & 0.008 & 0.004 & NM\\
   King 23 & 3109989392447848960 & 110.440528 & -0.960479 & 12.263 & 97.7 & 3 & 8100 & 13/12/2018 & 53.67 & 0.008 & 0.006 & \\
   King 23 & 3109989121872110080 & 110.46442 & -0.981852 & 12.373 & 83.8 & 3 & 8100 & 14/12/2018 & 53.846 & 0.019 & 0.005 & \\
   NGC\,581 & 509862169090128000 & 23.371715 & 60.646584 & 7.437 & 84.9 & 4 & 260 & 14/12/2018 & -46.782 & 0.008 & 0.021 & Variable/M supergiant\\
   NGC 1027 & 465682692367536384 & 40.577551 & 61.737139 & 8.077 & 82.4 & 3 & 360 & 16/12/2018 & -43.317 & 0.017 & 0.012 & \\
   NGC 1027 & 465853425905007872 & 39.571876 & 61.578574 & 10.138 & 84.8 & 3 & 2600 & 16/12/2018 & -2.322 & 0.013 & 0.013 & NM\\
   NGC 1647 & 3409869064229856128 & 71.521245 & 18.800733 & 6.91 & 197.0 & 3 & 360 & 16/12/2018 & -6.351 & 0.008 & 0.01 & \\
   NGC 1647 & 3410117313342257664 & 71.649563 & 19.494243 & 7.784 & 118.8 & 3 & 360 & 16/12/2018 & -6.483 & 0.013 & 0.042 & \\
   NGC 1750 & 3418663507987229184 & 76.087212 & 23.527111 & 6.892 & 191.0 & 3 & 360 & 14/12/2018 & -10.065 & 0.009 & 0.008 & \\
   NGC 1750 & 3418709412597592576 & 76.200009 & 23.920917 & 7.205 & 182.4 & 3 & 360 & 14/12/2018 & -13.481 & 0.014 & 0.018 & SB?/NM\\
   NGC 2186 & 3318650315417696384 & 93.026044 & 5.464272 & 9.486 & 137.4 & 3 & 2400 & 16/12/2018 & 20.821 & 0.017 & 0.071 & \\
   NGC 2281 & 951479674341906560 & 102.062876 & 41.072802 & 8.69 & 80.3 & 3 & 160 & 14/12/2018 & 19.552 & 0.016 & 0.018 & \\
   NGC 2281 & 951676899239237632 & 102.090499 & 41.302284 & 6.857 & 158.8 & 3 & 120 & 14/12/2018 & 19.407 & 0.009 & 0.005 & \\
   NGC 2345 & 3044669232011557760 & 107.109659 & -13.187338 & 9.891 & 71.8 & 3 & 1100 & 15/12/2018 & 58.598 & 0.027 & 0.017 & \\
   NGC 2345 & 3044669506889464320 & 107.091051 & -13.173124 & 9.234 & 78.2 & 3 & 1400 & 15/12/2018 & 64.078 & 0.027 & 0.118 & SB\\
   NGC 2345 & 3044665967836430976 & 107.126537 & -13.231258 & 9.706 & 91.1 & 3 & 2300 & 15/12/2018 & 58.384 & 0.01 & 0.032 & \\
   NGC 2358 & 2934838637556373888 & 109.256039 & -17.151366 & 9.634 & 132.4 & 3 & 2400 & 15/12/2018 & 28.496 & 0.006 & 0.005 & \\
   NGC 2358 & 2934861658580860928 & 109.314538 & -16.888067 & 9.117 & 226.8 & 3 & 2400 & 15/12/2018 & 27.003 & 0.015 & 0.008 & \\
   NGC 2358 & 2934834411310429824 & 109.317203 & -17.221546 & 9.992 & 160.2 & 3 & 2400 & 15/12/2018 & 28.56 & 0.019 & 0.048 & \\
   NGC\,7654 & 2015645057006454144 & 351.186937 & 61.344091 & 8.202 & 81.9 & 6 & 910 & 15/12/2018 & -32.012 & 0.014 & 0.021 & \\
   NGC\,7654 & 2015663649928489856 & 351.066006 & 61.588202 & 7.855 & 126.6 & 3 & 900 & 15/12/2018 & -32.199 & 0.009 & 0.057 & \\
   Sotck\,1 & 2021217798603195264 & 294.47485 & 24.712883 & 7.774 & 102.2 & 1 & 180 & 19/04/2019 & 19.605 & & 0.238 & \\
   Trumpler\,2 & 457691785453711360 & 39.169356 & 56.196734 & 6.814 & 194.2 & 3 & 360 & 15/12/2018 & -3.781 & 0.001 & 0.012 & \\
   Trumpler\,2 & 454676374812927744 & 39.220012 & 55.915377 & 6.783 & 191.5 & 3 & 360 & 15/12/2018 & -3.881 & 0.012 & 0.007 & \\
   UBC\,54 & 233049705784504576 & 64.672277 & 46.209049 & 9.799 & 114.5 & 3 & 2400 & 15/12/2018 & -15.278 & 0.017 & 0.02 & \\
   UBC\,54 & 233820772674404736 & 64.714494 & 46.561861 & 9.921 & 110.2 & 3 & 2400 & 16/12/2018 & -15.138 & 0.005 & 0.025 & \\
   \hline
   \end{tabular}\tablefoot{
(NM) nom-member; (SB) spectroscopic binary.
}
\label{tab:obs_log}
\end{table*}

\begin{table*}[]
    \centering
    \caption{Effective temperatures, surface gravities, and global metallicities obtained from the spectroscopic analysis of the observed stars. The CDS version includes all the individual abundances.}\label{tab:stars_info}
    \begin{tabular}{llcccc}
      \hline
      cluster &  star & $T_{\mathrm{eff}}$ [K] & $\log g$ & [M/H] &  [Fe/H]  \\
       \hline
          ASCC\,23 &   968512002808979456 &  $5022 \pm 18$ & $2.24 \pm 0.05$ & $-0.05$ & $-0.02 \pm 0.06$ \\
        Alessi\,44 &  4240034050119188352 &  $4667 \pm 10$ & $1.69 \pm 0.05$ & $-0.08$ & $ 0.04 \pm 0.05$ \\
        Alessi\,62 &  4519333643079413760 &  $5027 \pm 13$ & $2.76 \pm 0.03$ & $-0.02$ & $ 0.07 \pm 0.05$ \\
        Alessi\,62 &  4519335601584434048 &  $5032 \pm 14$ & $2.65 \pm 0.03$ & $ 0.03$ & $ 0.13 \pm 0.05$ \\
      COIN-Gaia\,2 &   423486184872160128 &  $5133 \pm 16$ & $2.72 \pm 0.03$ & $ 0.00$ & $ 0.05 \pm 0.05$ \\
      COIN-Gaia\,2 &   423487726755535232 &  $5144 \pm 21$ & $2.71 \pm 0.06$ & $-0.03$ & $ 0.02 \pm 0.05$ \\
      COIN-Gaia\,6 &   506223988186933888 &  $3929 \pm 12$ & $0.27 \pm 0.05$ & $-0.50$ &  \\
      COIN-Gaia\,6 &   506214401820077056 &  $5702 \pm 30$ & $2.61 \pm 0.07$ & $0.76$ & \\
      COIN-Gaia\,6 &   506224812820610176 &  $5489 \pm 31$ & $2.18 \pm 0.07$ & $0.58$ & \\
         FSR\,0951 &  3344692140944457088 &  $5152 \pm 13$ & $2.63 \pm 0.04$ & $-0.06$ & $ 0.00 \pm 0.06$ \\
         FSR\,0951 &  3344673277448107776 &  $5180 \pm 14$ & $2.60 \pm 0.03$ & $-0.07$ & $-0.02 \pm 0.05$ \\
          King\,23 &  3109989392447848960 &  $4492 \pm 15$ & $2.07 \pm 0.03$ & $-0.10$ & $-0.07 \pm 0.07$ \\
          King\,23 &  3109989121872110080 &  $4619 \pm 15$ & $2.21 \pm 0.03$ & $-0.07$ & $-0.02 \pm 0.08$ \\
         NGC\,1027 &   465682692367536384 &  $4160 \pm 14$ & $1.07 \pm 0.03$ & $-0.23$ & $-0.21 \pm 0.10$ \\
         NGC\,1647 &  3410117313341005184 &  $4696 \pm 11$ & $1.81 \pm 0.05$ & $-0.16$ & $-0.08 \pm 0.07$ \\
         NGC\,1647 &  3409869064229856128 &  $4914 \pm  9$ & $1.80 \pm 0.02$ & $-0.13$ & $-0.04 \pm 0.05$ \\
         NGC\,1750 &  3418663507987229184 &  $4128 \pm  8$ & $1.10 \pm 0.02$ & $-0.32$ & $-0.26 \pm 0.06$ \\
         NGC\,2186 &  3318650315417696384 &  $5686 \pm 14$ & $1.72 \pm 0.04$ & $-0.05$ & $ 0.04 \pm 0.06$ \\
         NGC\,2281 &   951676899239237632 &  $4451 \pm  8$ & $1.74 \pm 0.03$ & $-0.18$ & $-0.11 \pm 0.04$ \\
         NGC\,2281 &   951479674341906560 &  $5199 \pm 12$ & $2.81 \pm 0.04$ & $ 0.11$ & $ 0.15 \pm 0.08$ \\
         NGC\,2345 &  3044665967836430976 &  $4251 \pm 14$ & $1.08 \pm 0.04$ & $-0.26$ & $-0.23 \pm 0.09$ \\
         NGC\,2345 &  3044669232011557760 &  $4468 \pm 13$ & $1.28 \pm 0.05$ & $-0.16$ & $-0.11 \pm 0.09$ \\
         NGC\,2358 &  2934861658580860928 &  $5044 \pm 11$ & $2.57 \pm 0.02$ & $-0.09$ & $-0.02 \pm 0.05$ \\
         NGC\,2358 &  2934838637556373888 &  $5129 \pm 13$ & $2.62 \pm 0.03$ & $-0.04$ & $ 0.03 \pm 0.05$ \\
         NGC\,2358 &  2934834411308570624 &  $5155 \pm 11$ & $2.88 \pm 0.03$ & $-0.07$ & $ 0.00 \pm 0.04$ \\
         NGC\,7654 &  2015645057006454144 &  $4001 \pm 12$ & $0.18 \pm 0.04$ & $-0.38$ &  \\
         NGC\,7654 &  2015663649928489856 &  $6072 \pm 16$ & $1.16 \pm 0.03$ & $-0.19$ & $-0.05 \pm 0.09$ \\
          Stock\,1 &  2024990807120270976 &  $4996 \pm 17$ & $2.58 \pm 0.03$ & $ 0.06$ & $ 0.15 \pm 0.07$ \\
       Trumpler\,2 &   454676374812927744 &  $4071 \pm  7$ & $0.83 \pm 0.03$ & $-0.40$ & $-0.34 \pm 0.05$ \\
       Trumpler\,2 &   457691785453711360 &  $4370 \pm  8$ & $1.30 \pm 0.03$ & $-0.26$ & $-0.19 \pm 0.04$ \\
           UBC\,54 &   233820772674404736 &  $5064 \pm 19$ & $2.47 \pm 0.03$ & $-0.04$ & $ 0.05 \pm 0.05$ \\
           UBC\,54 &   233049705784504576 &  $5089 \pm 14$ & $2.27 \pm 0.04$ & $-0.05$ & $ 0.01 \pm 0.05$ \\
       \hline
      \end{tabular}    \tablefoot{We also list the iron abundances computed with respect to the Sun ([Fe/H]). The stars without abundance values correspond to the two identified supergiants (COIN-Gaia 6 and NGC 7654), and the two stars with low S/N are from COIN-Gaia 6, as discussed in Sect~\ref{sect:atmos_dete}.
}
\end{table*}

\begin{table*}
\setlength{\tabcolsep}{0.5mm}
\begin{center}
\caption{Comparison of the atmospheres parameters determined by different authors.}
\begin{tabular}{lccccccccccc}
\hline
Star & \multicolumn{4}{l}{\teff [K]} & \multicolumn{4}{l}{\logg} &\multicolumn{3}{l}{[Fe/H]}\\
 & Here & R16 & A19 & H19 & Here & R16 & A19 & H19  & Here & A19 & H19\\
\hline
3044665967836430976 & 4251$\pm$14 & 4300 & 4183$\pm$52 & 4020 & 1.08$\pm$0.04 & 1.60 & 0.96$\pm$0.09 & 1.03 & -0.22$\pm$0.09 & -0.39$\pm$0.08 & -0.28$\pm$0.07\\
3044669232011557760 & 4468$\pm$13 & 4300 & 4283$\pm$25 & 4350 & 1.28$\pm$0.05 & 1.20 & 1.06$\pm$0.09 & 1.60 & -0.11$\pm$0.09 & -0.29$\pm$0.04 & -0.32$\pm$0.07\\
\hline
\end{tabular}
\tablebib{
(R16)~\citet{Reddy+2016}; (A19) \citet{alonso_santiago2019}; (H19) \citet{holanda2019}.
}
\label{tab:ngc2345_lit}
\end{center}
\end{table*}

\begin{table*}
\setlength{\tabcolsep}{0.5mm}
\begin{center}
\caption{Obtained solar abundances.}
\begin{tabular}{llrr}
   \hline
    element &  $A(X)$ \\
   \hline
    \mbox{Al\,{\sc i}} &  $6.46 \pm 0.01$ \\
    \mbox{Ba\,{\sc ii}} &  $2.16 \pm 0.01$ \\
    \mbox{Ca\,{\sc i}} &  $6.37 \pm 0.04$ \\
    \mbox{Ce\,{\sc ii}} &  $1.40 \pm 0.01$ \\
    \mbox{Co\,{\sc i}} &  $4.82 \pm 0.03$ \\
    \mbox{Cr\,{\sc i}} &  $5.60 \pm 0.03$ \\
    \mbox{Fe\,{\sc i}} &  $7.43 \pm 0.04$ \\
    \mbox{Fe\,{\sc ii}} &  $7.42 \pm 0.05$ \\
    \mbox{Mg\,{\sc i}} &  $7.52 \pm 0.01$ \\
    \mbox{Mn\,{\sc i}} &  $5.44 \pm 0.04$ \\
    \mbox{Na\,{\sc i}} &  $6.32 \pm 0.01$ \\
    \mbox{Nd\,{\sc ii}} &  $1.33 \pm 0.01$ \\
    \mbox{Ni\,{\sc i}} &  $6.18 \pm 0.04$ \\
    \mbox{Sc\,{\sc ii}} &  $3.15 \pm 0.03$ \\
    \mbox{Si\,{\sc i}} &  $7.51 \pm 0.03$ \\
    \mbox{Ti\,{\sc i}} &  $4.86 \pm 0.04$ \\
    \mbox{Ti\,{\sc ii}} &  $4.88 \pm 0.04$ \\
    \mbox{V\,{\sc i}} &  $3.86 \pm 0.03$ \\
    \mbox{Y\,{\sc ii}} &  $2.16 \pm 0.01$ \\
   \hline
   \end{tabular}\label{tab:solar_abu}
\end{center}
\end{table*}

\begin{table*}
\setlength{\tabcolsep}{0.5mm}
\begin{center}
\caption{Average abundances for the observed clusters. We included only a few elements. The CDS version includes all the elements studied.}
\begin{tabular}{lcccccc}
   \hline
   Cluster & [Mg/H] & [Si/H] & [Ca/H] & [Fe/H] & [Ba/H] & N \\
   \hline
     ASCC\,23 & -0.04$\pm$0.04 & 0.14$\pm$0.08 & 0.05$\pm$0.07 & -0.02$\pm$0.06 & 0.45$\pm$0.11 & 1\\
     Alessi\,44 & 0.00$\pm$0.02 & 0.13$\pm$0.05 & 0.05$\pm$0.03 & 0.04$\pm$0.05 & 0.19$\pm$0.09 & 1\\
     Alessi\,62 & 0.09$\pm$0.0 & 0.12$\pm$0.07 & 0.16$\pm$0.02 & 0.10$\pm$0.04 & 0.31$\pm$0.01 & 2\\
     COIN-Gaia\,2 & 0.01$\pm$0.01 & 0.01$\pm$0.01 & 0.06$\pm$0.01 & 0.04$\pm$0.02 & 0.32$\pm$0.11 & 2\\
   \hline
   \end{tabular}\label{tab:avg_abu}
\end{center}
\end{table*}

\begin{table*}
\begin{center}
\caption{Orbital parameters obtained for the studied clusters}
\begin{tabular}{lcccccc}
   \hline
     Cluster & $r_{apogeo}$ & $r_{perigeo}$ & $R_{birth}$ & eccentricity & $z_{max}$ & phi\\
   \hline
     ASCC\,23 & 9.48$\pm$0.04 & 7.75$\pm$0.09 & 8.28$\pm$0.03 & 0.1$\pm$0.004 & 0.22$\pm$0.01 & 5.29$\pm$2.2\\
     Alessi\,44 & 8.35$\pm$0.06 & 6.82$\pm$0.05 & 8.19$\pm$0.12 & 0.1$\pm$0.007 & 0.11$\pm$0.01 & 5.75$\pm$0.04\\
     Alessi\,62 & 9.16$\pm$0.04 & 6.34$\pm$0.04 & 9.04$\pm$0.02 & 0.18$\pm$0.001 & 0.17$\pm$0.01 & 4.74$\pm$0.05\\
     COIN-Gaia\,2 & 9.5$\pm$0.08 & 7.8$\pm$0.14 & 8.07$\pm$0.22 & 0.1$\pm$0.011 & 0.14$\pm$0.01 & 1.24$\pm$0.12\\
     COIN-Gaia\,6 & 13.35$\pm$0.47 & 10.42$\pm$0.14 & 13.15$\pm$0.45 & 0.12$\pm$0.013 & 0.26$\pm$0.04 & 3.36$\pm$0.47\\
     FSR\,0951 & 10.32$\pm$0.1 & 7.79$\pm$0.22 & 9.57$\pm$0.09 & 0.14$\pm$0.016 & 0.17$\pm$0.02 & 5.66$\pm$0.57\\
     King\,23 & 10.91$\pm$0.25 & 7.47$\pm$0.25 & 9.03$\pm$0.98 & 0.19$\pm$0.012 & 0.37$\pm$0.02 & 2.46$\pm$1.37\\
     NGC\,581 & 9.81$\pm$0.08 & 9.7$\pm$0.07 & 9.7$\pm$0.07 & 0.01$\pm$0.001 & 0.06$\pm$0.01 & 5.88$\pm$0.01\\
     NGC\,1027 & 9.58$\pm$0.06 & 7.35$\pm$0.18 & 8.48$\pm$0.06 & 0.13$\pm$0.01 & 0.21$\pm$0.01 & 0.97$\pm$0.10\\
     NGC\,1647 & 9.7$\pm$0.05 & 8.01$\pm$0.07 & 8.65$\pm$0.07 & 0.1$\pm$0.003 & 0.18$\pm$0.01 & 2.77$\pm$0.05\\
     NGC\,1750 & 9.68$\pm$0.05 & 8.08$\pm$0.09 & 8.48$\pm$0.1 & 0.09$\pm$0.004 & 0.12$\pm$0.01 & 5.85$\pm$0.09\\
     NGC\,2186 & 10.31$\pm$0.11 & 8.65$\pm$0.1 & 9.14$\pm$0.1 & 0.09$\pm$0.007 & 0.22$\pm$0.01 & 0.46$\pm$0.06\\
     NGC\,2281 & 8.58$\pm$0.06 & 7.26$\pm$0.01 & 8.44$\pm$0.04 & 0.08$\pm$0.004 & 0.18$\pm$0.01 & 0.91$\pm$0.04\\
     NGC\,2345 & 10.01$\pm$0.11 & 9.16$\pm$0.17 & 9.39$\pm$0.21 & 0.04$\pm$0.005 & 0.09$\pm$0.01 & 1.28$\pm$0.08\\
     NGC\,2358 & 8.86$\pm$0.03 & 7.08$\pm$0.06 & 8.69$\pm$0.05 & 0.11$\pm$0.003 & 0.03$\pm$0.01 & 5.59$\pm$0.06\\
     NGC\,7654 & 8.79$\pm$0.04 & 7.69$\pm$0.13 & 7.69$\pm$0.12 & 0.07$\pm$0.006 & 0.06$\pm$0.01 & 2.07$\pm$0.08\\
     Stock\,1 & 8.19$\pm$0.02 & 6.5$\pm$0.09 & 7.95$\pm$0.13 & 0.12$\pm$0.007 & 0.06$\pm$0.01 & 5.49$\pm$0.05\\
     Trumpler\,2 & 8.54$\pm$0.03 & 8.08$\pm$0.05 & 8.08$\pm$0.05 & 0.03$\pm$0.003 & 0.1$\pm$0.01 & 3.26$\pm$0.02\\
     UBC\,54 & 9.02$\pm$0.05 & 7.12$\pm$0.05 & 7.68$\pm$0.04 & 0.12$\pm$0.004 & 0.1$\pm$0.01 & 5.11$\pm$0.02\\
   \hline
   \end{tabular}
   \label{tab:orb_par}
\end{center}
\end{table*}

\end{document}